\documentclass[10pt]{article}

\usepackage{amsmath}

\usepackage{amssymb}

\usepackage{url}

\usepackage{graphicx}

\usepackage{cite}

\usepackage{color}

\usepackage[labelfont=bf,labelsep=period,justification=raggedright]{caption}

\makeatletter

\renewcommand{\@biblabel}[1]{\quad#1.}

\makeatother

\date{}

\begin{document}


\begin{flushleft} {\Large \textbf{On the accuracy of language trees} }
\vspace{0.5cm}
\\
Simone Pompei$^{1,2}$, Vittorio Loreto$^{1,3}$, Francesca Tria$^{1,\ast}$
\vspace{0.5cm}
\\
\bf{1} Complex Systems Lagrange Lab, Institute for
Scientific Interchange (ISI), Via S. Severo 65, 10133, Torino, Italy
\\
\bf{2} Universit\`a di Torino, Physics Dept., Via Giuria 2, 10125, Torino, Italy
\\
\bf{3} Sapienza Universit\`a di Roma, Physics Dept., Piazzale  Aldo Moro 2, 00185, Rome, Italy
\\
\vspace{0.5cm}
$\ast$ E-mail: fra\_trig@yahoo.it \end{flushleft}


\section*{Abstract}

Historical linguistics aims at inferring the most likely language phylogenetic tree starting from information concerning the evolutionary relatedness of languages. The available information are typically lists of homologous (lexical, phonological, syntactic) features or characters for many different languages: a set of parallel corpora whose compilation represents a paramount achievement in linguistics.

From this perspective the reconstruction of language trees is an example of inverse problems: starting from present, incomplete and often noisy, information, one aims at inferring the most likely past evolutionary history.  A fundamental issue in inverse problems is the evaluation of the inference made. A standard way of dealing with this question is to generate data with artificial models in order to have full access to the evolutionary process one is going to infer.  This procedure presents an intrinsic limitation: when dealing with real data sets, one typically does not know which model of evolution is the most suitable for them. A possible way out is to compare algorithmic inference with expert classifications.  This is the point of view we take here by conducting a thorough survey of the accuracy of reconstruction methods as compared with the Ethnologue expert classifications. We focus in particular on state-of-the-art {\it distance-based} methods for phylogeny reconstruction using worldwide linguistic databases.

In order to assess the accuracy of the inferred trees we introduce and characterize two generalizations of standard definitions of distances between trees.  Based on these scores we quantify the relative performances of the distance-based algorithms considered. Further we quantify how the completeness and the coverage of the available databases affect the accuracy of the reconstruction. Finally we draw some conclusions about where the accuracy of the reconstructions in historical linguistics stands and about the leading directions to improve it.

\section*{Introduction}

The last few years have seen a wave of computational approaches devoted to historical linguistics~\cite{Renfrew_2000,handbook_2004,diachronica_2010}, mainly centred around phylogenetic methods. While the first aim of phylogeny reconstruction is that of classifying a set of species (viruses, biological species, languages, texts), the information embodied in the inferred trees goes beyond a simple classification knowledge. Statistical tools~\cite{stat_edge_1,stat_edge_2, stat_edge_3, stat_edge_4,stat_edge_5,stat_edge_6}, for instance, permit to assign time weights to the edges of a phylogenetic tree, giving the opportunity to gather information about the past history of the whole evolutionary process. These techniques have been successfully employed to investigate features of human prehistory~\cite{Gray_2003,Bryant_2005,Pagel_2007,Atkinson_2008,Dunn_2008,Gray_2009}.

The application of computational tools in historical linguistics is not a novel one, since it dates back to the 50's, when Swadesh~\cite{swadesh_1952,swadesh_1955} first proposed an approach to comparative linguistics that involved the quantitative comparison of lexical cognates, an approach named {\em lexicostatistics}. The most important element here is the compilation, for each language being considered, of lists of universally used meanings (hand, mouth, sky, I, ..). The initial set of meanings included $200$ items which were then reduced down to $100$, including some new terms which were not in his original list. Each language is represented by its specific list and different languages can be compared exploiting the similarity of their lists. The similarity is assessed by estimating the level of cognacy between pairs of words. The higher the proportion of cognacy the closer the languages are related. Though originally cognacy decisions was solely based on the work of trained and experienced linguists, automated methods have been progressively introduced (see~\cite{nerbonne_99} and for a recent overview ~\cite{LDN_comp}) that exploit the notion of {\em Edit Distance} (or {\em Levenshtein Distance})~\cite{Lev_1966} between words, considered as strings of characters. The computation of the Edit Distance between all the pairs of homologous words in pairs of languages leads to the computation of a ``distance'' between pairs of languages. This value is entered into a $N \times N$ table of distances, where $N$ is the number of languages being compared. This distance matrix can then be submitted to {\it distance-based} algorithms for the purpose of generating trees showing relationships among languages.

The construction of the distance matrix is of course a crucial step since the reliability of the reconstruction of the evolutionary history, i.e., the outcome of a phylogenetic reconstruction method, strongly depends on the properties of the distance matrix. In particular if the matrix features the property of being {\it additive}, there are algorithms that guarantee the reconstruction of the unique true tree (see~\cite{M3AS} for a recent overview). A distance matrix is said to be additive if it can be constructed as the sum of a tree's branch lengths. When considering experimental data, additivity is almost always violated. Violations of additivity can arise both from experimental noise and from properties of the evolutionary process underlying the data. One of the possible sources of violation of additivity is the so-called back-mutation: in old phylogenies a single character can experience multiple mutations. In this case the distances between taxa are no longer proportional to their evolutionary distances. In historical linguistics this would happen if one was considering meanings that change very rapidly. For this reason linguists are typically interested in removing from the lists all the fast-evolving meanings. Of course this is not an easy task, bringing inextricably with itself a fair amount of arbitrariness in the choice. Along the same lines another crucial difficulty in lexicostatistics concerns the rate of change of the individual meanings. Different meanings, represented in each language by different words, evolve with different rates of change. In a biological parallel one would say that the mutation rate, i.e., the rate over which specific words undergo morphological, phonetic or semantic changes, are meaning dependent. This effect again is not easily cured and again different choices of the list composition could lead to different reconstructions. Finally another source of deviations from additivity is the so-called horizontal-transfer. The reconstruction of a phylogeny from data underlies the assumption that information flows vertically from ancestors to offspring. However, in many processes information also flows horizontally. In historical linguistics borrowings represent a well-known confounding factor for a correct phylogenetic inference.

All the fore-mentioned difficulties in the reconstruction of phylogenetic trees strongly call for reliable methods to evaluate the reconstructed phylogenies. Along with this it comes the need of valid benchmarks for determining the reliability of the different methods used to reconstruct phylogenetic trees. The standard way of testing the proposed algorithms is the construction of models to generate artificial phylogenies~\cite{M3AS, sbix_2010, Fast-SBiX}, so that the algorithmic results can be directly compared with the true, known, observables of interest. However, in doing that, one makes inevitable assumptions on the evolutionary processes of interest, which can in turn influence the reconstruction performance.  To overcome this problem, we consider here an application of phylogenetic tools to historical linguistics. This field offers a good reference point, since classifications made with phylogenetic tools can be compared with catalogues of languages made by experts. We focus in particular on the Ethnologue classification. The Ethnologue can be described as a comprehensive catalogue of the known languages spoken in the world~\cite{ethnologue}, organized by continent and country, being thus a valid reference point to evaluate trees inferred using phylogenetic algorithms (see section {\it Data} for details).

Here we evaluate trees reconstructed using {\em distance-based} phylogenetic methods against the Ethnologue trees. To this end it is important to set the tools to compare expert Ethnologue trees and phylogenetically inferred trees.  There are several standard ways of measuring the distance between two phylogenetic trees. Here we take into account two of them, the Robinson-Foulds (RF) distance~\cite{Robinson_Foulds_1981}, which counts the number of bipartitions on which the two trees differ, and the Quartet Distance (QD)~\cite{QD}, which counts the number of subset of four taxa on which the two trees differ.

A technical problem when comparing Ethnologue classifications and inferred trees is that typically Ethnologue trees are not binary while all the inferred trees are. In order to overcome this difficulty we introduce two incompatibilities scores, which are two generalizations of both the Robinson-Foulds~\cite{Robinson_Foulds_1981} and the Quartet Distance measures~\cite {QD}. We present results obtained on a wide range of language families. This allows to compare different definitions of distances as well as different reconstruction algorithms.

The outline of the paper is as follows. We first introduce the {\em Ethnologue}~\cite{ethnologue} project and both the {\em Automated Similarity Judgement Program} ({\em ASJP})\cite{ASJP} and the {\em Austronesian Basic Vocabulary Database} ({\em ABVD})~\cite{ABVD} database we used in our analysis, pointing out some structural and statistical features that will be relevant in our discussion.  Next we introduce some mathematical tools. We define both the {\em Levenshtein Normalized Distance} ({\em LDN}) and the {\em Levenshtein Divided Normalized Distance}({\em LDND})~\cite{LDN_comp} to compute a ``distance'' between lists of word. The quantification of the accuracy of the inference of language trees we present is achieved with the Robinson-Foulds distance (RF)~\cite{Robinson_Foulds_1981} and the Quartet Distance (QD)~\cite {QD}. These are two standard definitions of distances between trees. We introduce and characterize such mathematical tools and we also present generalizations of these two scores, in order to adapt them for the comparison of binary (inferred) and non-binary (classifications) trees.  We then present the results of the comparisons between the Ethnologue classifications and the language trees inferred based on the ASJP database. We first consider the ASJP database in order to perform a worldwide, i.e., large-scale, analysis. Finally we point out how some of the properties of word-lists, such as the completeness and the coverage, affect the accuracy of the reconstruction. To this end we present a comparative analysis on the inference of the Austronesian family, making use of both the ASJP and the ABVD database. The Supporting Information provides an extensive account of the whole set of results we obtained.

\section*{Materials and Methods}

\subsection*{Data} 

\label{data}

\begin{figure}[htp] \begin{center} \includegraphics[width=1.05\textwidth]{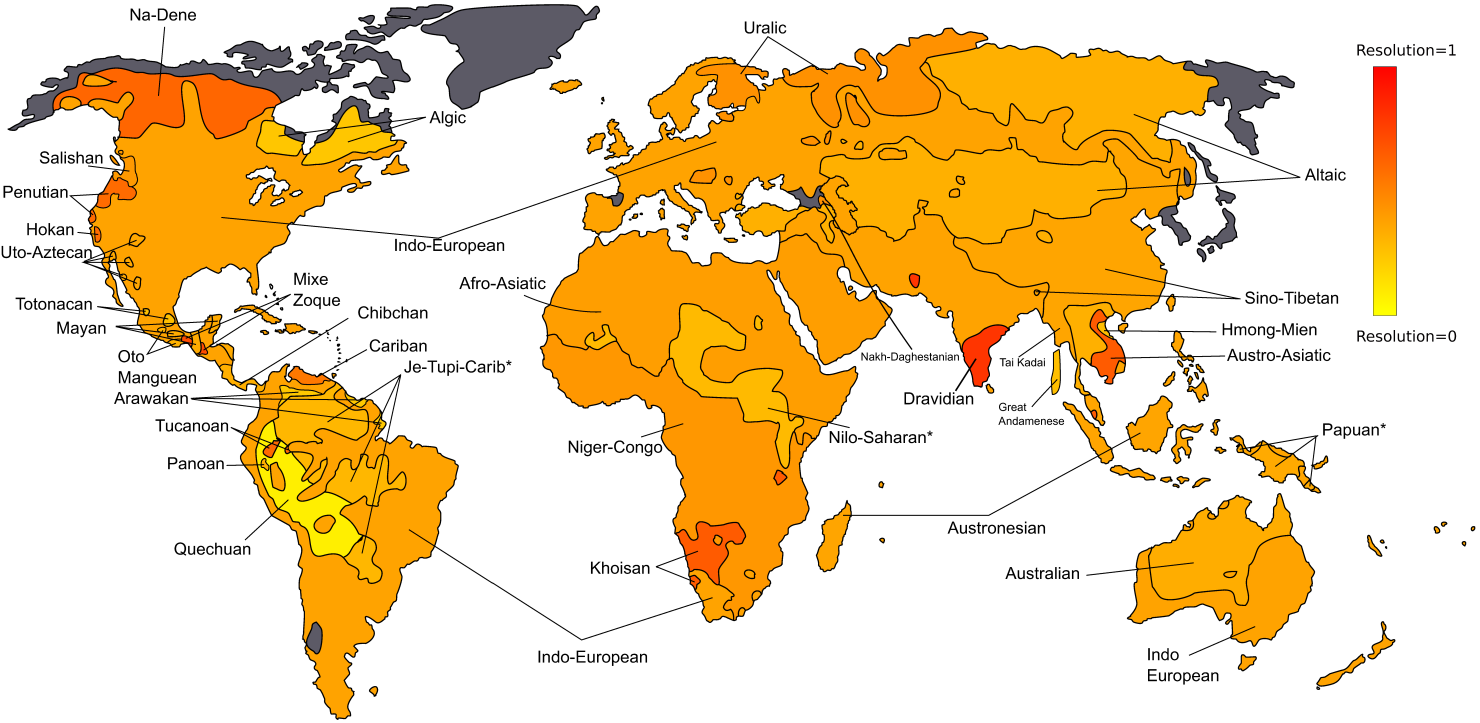} 

\end{center} \caption{ {\bf Ethnologue resolution power.} This map represents the Ethnologue resolution power in the different world locations. Red areas corresponds to regions where the Ethnologue classification is completely binary, i.e., correspond to a tree in which each internal node has exactly two child nodes. Yellow areas corresponds to fully unspecified trees, featuring only a star structure. Grey areas are those for which no data are present in the databases we consider to reconstruct language trees. Asterisks are for regions which include more than one language family (we report in the Supporting Information  the list of such families).}  \label{ethnologue_binary} \end{figure}

The {\bf Ethnologue} can be described as a comprehensive catalogue of the known languages spoken in the world~\cite{ethnologue}. The Ethnologue was founded by R.S. Pittman in 1951 as a way to communicate with colleagues about language development projects. Its first edition was a ten-page informal list of $46$ language and language group names. As of its sixteenth edition, Ethnologue has grown into a comprehensive database that is constantly being updated as new information arrives. As of now it contains close to $7000$ language descriptions, organized by continent and country, which can be represented as a tree. As already mentioned, this tree is not always fully specified since it contains a lot of non-binary structures,  in which the details of the phylogeny are not given due to a lack of certain information. Figure~\ref{ethnologue_binary} illustrates geographically how the Ethnologue classifications deviate from being purely binary.

\begin{figure}[htp]

\begin{center}

  \includegraphics[width=\textwidth]{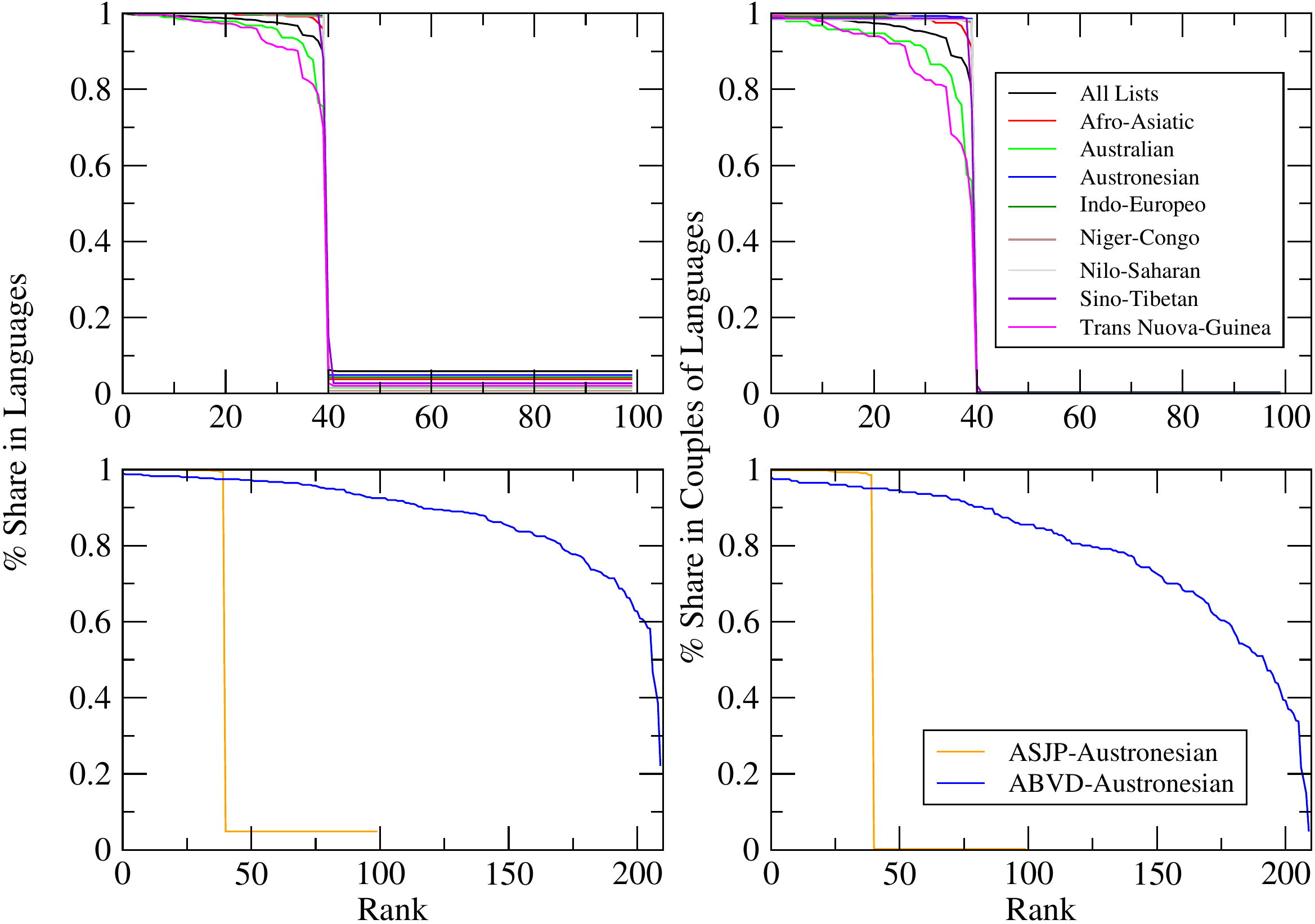} \end{center} \caption{Top: {\bf Statistics of the ASJP database.} {\bf (left panel)} Fraction-rank plot: for each word in the lists of words of the Automated Similarity Judgement Project (ASJP), we measured the fraction of languages containing it. The plot reports this fraction vs. its rank.  In the $100$-items lists in the ASJP database, only $40$ meanings are shared by almost $100\%$ of the languages for each family. {\bf (right panel)} Ranked fraction of pairs of languages sharing each specific word vs. rank. Again only $40$ meanings are shared by almost $100\%$ of the pairs of languages. Bottom: {\bf Statistical measures on the ABVD database.} {\bf (left panel)} Fraction-rank plot: for each word in the lists of words of the Austronesian Basic Vocabulary Database (ABVD), we measured the fraction of languages containing it. The plot reports this fraction vs. its rank. {\bf (right panel)} Ranked fraction of pairs of languages sharing each specific word vs. rank. For sake of a rough comparison we also reported the same quantities measured on the Austronesian family of the ASJP database. The ASJP includes $40$ words up to a maximum of almost $100\%$ of the languages, whereas in the ABVD the percentage of coverage is at least of $50\%$ for almost all the words in the list. Limited to the $40$ most shared words the ASJP database features a slightly larger coverage than the ABVD database. }

\label{comparison}

\end{figure}

The {\bf Automated Similarity Judgement Program} ({\bf ASJP})\cite{ASJP} includes $100$-items word lists of about 50 families of languages throughout the world. These lists are written in a standardized orthography (ASJP code) which employs only symbols of the standard QWERTY keyboard, defining vowels, consonants and phonological features. The full database is available at \url{http://email.eva.mpg.de/~wichmann/ASJPHomePage.htm}. Figure~\ref{comparison} (top) reports two statistical measures on the database to quantify its completeness. In particular we report the ranked fraction of languages containing a word for a specific meaning vs. the rank (left panel) and the ranked fraction of pairs of languages sharing a word (not necessarily a cognate) for a specific meaning vs. the rank (right panel). The second measure helps in understanding how accurate, from a statistical point of view, the computation of the distance between two languages averaging the Levenshtein distances of all the words for homologous meanings. Evidently the database is very complete up to $40$ meanings.

The {\bf Austronesian Basic Vocabulary Database} ({\bf ABVD})~\cite{ABVD} contains lexical items from $737$ languages (as of January 2011) spoken throughout the Pacific region. Most of these languages belong to the Austronesian language family, which is the largest family in the world. Due to the extended and phonetic characters used for the lexical orthography, all the information is encoded in the Unicode format UTF-8. The web site of the database is \url{http://language.psy.auckland.ac.nz/austronesian/} and we downloaded it on October, the 4th 2010. We focused in particular on a subset of $305$ languages that are present both in the ASJP database and in the Ethnologue classification. Figure~\ref{comparison} (bottom) reports the same quantities of Figure~\ref{comparison} (top) for the ABVD database. It is evident how, limited to the Austronesian family, the ABVD database features an overall larger (with respect to the ASJP database) number of meanings across all the languages considered. The level of coverage decreases progressively as one increases the number of meanings. A word of caution is in order. It is of course not possible to compare the completeness of the ASJP and the ABVD databases since they refer to two completely different projects with different aims: ASJP aiming at a full coverage of the Swadesh lists on all the world languages and ABVD being focused only on the Austronesian languages. It is nevertheless interesting to compare them only as for the Austronesian family is concerned. We shall come back on this point when we shall compare the accuracy of the reconstructed trees using different databases.

\subsection*{Distance between languages}

In our studies we represent a language by its list of words for the different meanings. The distance between two languages is based on the distance between pairs of words corresponding to homologous meanings in the two lists. The distance between two words is computed by means of the Levenshtein distance (LD). The LD is a metric to quantify the difference between two sequences and it is defined as the minimum number of edit operations needed to transform one string into the other, the allowable edit operations being insertion of a character, deletion of a character and substitution of a single character.

Once the distance between pairs of words is specified, two different definitions of distances between languages have been introduced~\cite{LDN_1,LDN_2,LDND,LDN_comp}: the {\it Levenshtein Distance Normalized (LDN)} and a revised interpretation of it named {\it Levenshtein Distance Normalized Divided (LDND)}. Both these definitions have been introduced to correctly define distances between languages, instead of simply considering an average of the LD distance between words corresponding to homologous meanings in the lists.

According to LDN definition~\cite{LDN_1,LDN_2}, given two words $\alpha_{i}$ and $\beta_{j}$, their distance is given by:

\begin{equation}
LDN(\alpha_{i},\beta_{j})=\frac{LD(\alpha_{i},\beta_{j})}{l(\alpha_{i},\beta_{j})}
\label{LDN1}
\end{equation}

\noindent where $LD(\alpha_{i},\beta_{j})$ is the LD between the two words and $l(\alpha_{i},\beta_{j})$ is the number of characters of the longest of the two words $\alpha_{i}$ and $\beta_{j}$ . This normalization has been introduced in order to avoid biases due to long words, giving in this way the same weight to all the words in the lists. Starting from this definition, let us now assume that the number of languages is $N$ and the list of meanings for each language contains $M$ items. Each language in the group is labelled by a Greek letter (say $\alpha$) and each word of that language by $\alpha_{i}$, with $1\leq i\leq M$. Then, two words $\alpha_{i}$ and $\beta_{j}$ in the languages $\alpha$ and $\beta$ have the same meaning (they correspond to the same meaning) if $i=j$ . The LDN between the two languages is thus:

\begin{equation} LDN(\alpha,\beta)=\frac{1}{M}\underset{i}{\sum}LDN(\alpha_{i},\beta_{i})
\label{LDN2}
\end{equation}

Another definition of distance between pair of languages has been introduced in~\cite{LDND} in order to avoid  biases due to accidental  orthographical similarities in  two languages:

\begin{equation} 
\Gamma(\alpha,\beta)=\frac{1}{M(M-1)}\underset{i\neq j}{\sum}LDN(\alpha_{i},\beta_{i})
\label{LDND_norm}
\end{equation}

\noindent The LDND distance between two languages is then defined as:

\begin{equation} LDND(\alpha,\beta)=\frac{LDN(\alpha,\beta)}{\Gamma(\alpha,\beta)}
\label{LDND}
\end{equation}

\noindent A comparison of the two definition of distances has been presented in~\cite{LDN_comp}. In the following we consider both these definitions of distances between languages; the dissimilarity-matrices computed according to them will be the starting point for the inference of the family trees, which will be compared with the corresponding Ethnologue classifications.

\subsection*{Robinson-Foulds, Quartet Distance and generalizations}

All the conclusions drawn in this work will be based on a quantitative comparison between inferred trees and the Ethnologue classifications. To this end it is important to recall how to measure the distance between two tree topologies. Here we recall in particular the mathematical definitions of two metrics between trees: the Robinson-Foulds distance (RF)~\cite{Robinson_Foulds_1981} and the Quartet Distance (QD)~\cite {QD}.

The Robinson-Foulds (RF) distance between two trees counts the number of bipartitions on which the two trees differ. If we delete an internal edge in a tree, the leaves will be divided in two subsets; we call this division a bipartition. Here we consider a normalized version of the RF distance, which counts the percentage of unshared bipartitions between two trees.  More formally, let ${\cal{T}}1$ and ${\cal{T}}2$ be two trees with the same set of leaves, then:

\begin{equation} 
RF({\cal{T}}1,{\cal{T}}2)=\frac{i({\cal{T}}1)+i({\cal{T}}2)-2e({\cal{T}}1,{\cal{T}}2)} {i({\cal{T}}1)+i({\cal{T}}2)} 
\label{RF}
\end{equation}

\noindent where $i({\cal{T}})$ denotes the set of internal edge of ${\cal{T}}$ and $e({\cal{T}}1,{\cal{T}}2)$ denotes the number of pairs of identical bipartitions in ${\cal{T}}1$ and ${\cal{T}}2$. The RF distance is a metric in the space of trees, whose value ranges from $0$ (if and only if ${\cal{T}}1={\cal{T}}2$ ) to $1$.

Another possible distance between two trees is the Quartet Distance (QD). In a tree of $N$ leaves, we can look at the subtrees defined by sets of four taxa (quartets).  In the general case of non fully resolved trees, a {\it butterfly} names a quartet in which the two pairs of leaves are divided by an internal edge and a {\it star} a quartet in which the leaves are all linked to the same node. The QD between two trees counts the number of non compatible quartets in the two trees. It is defined as:

\begin{equation}
QD({\cal{T}}1,{\cal{T}}2)=\frac{q({\cal{T}}1)+q({\cal{T}}2)-2s({\cal{T}}1,{\cal{T}}2)-d({\cal{T}}1,{\cal{T}}2)} {\mathrm{norm}(N)}
\label{QD}
\end{equation}

\noindent where $q({\cal{T}})$ is the total number of butterflies in ${\cal{T}}$, $s({\cal{T}}1,{\cal{T}}2)$ is the number of identical butterflies in ${\cal{T}}1$ and ${\cal{T}}2$ and $d({\cal{T}}1,{\cal{T}}2)$ is the number of different butterflies in the two trees. The normalization factor is the number, $\mathrm{norm}(N)=\binom{N}{4}$, of quartets in a tree of $N$ taxa.  The QD, as well as the RF distance, is a metric in the space of trees, whose value ranges from $0$ (if and only if ${\cal{T}}1={\cal{T}}2$ ) to $1$.

\begin{figure}[htp] \begin{center} \includegraphics[width=0.8\textwidth]{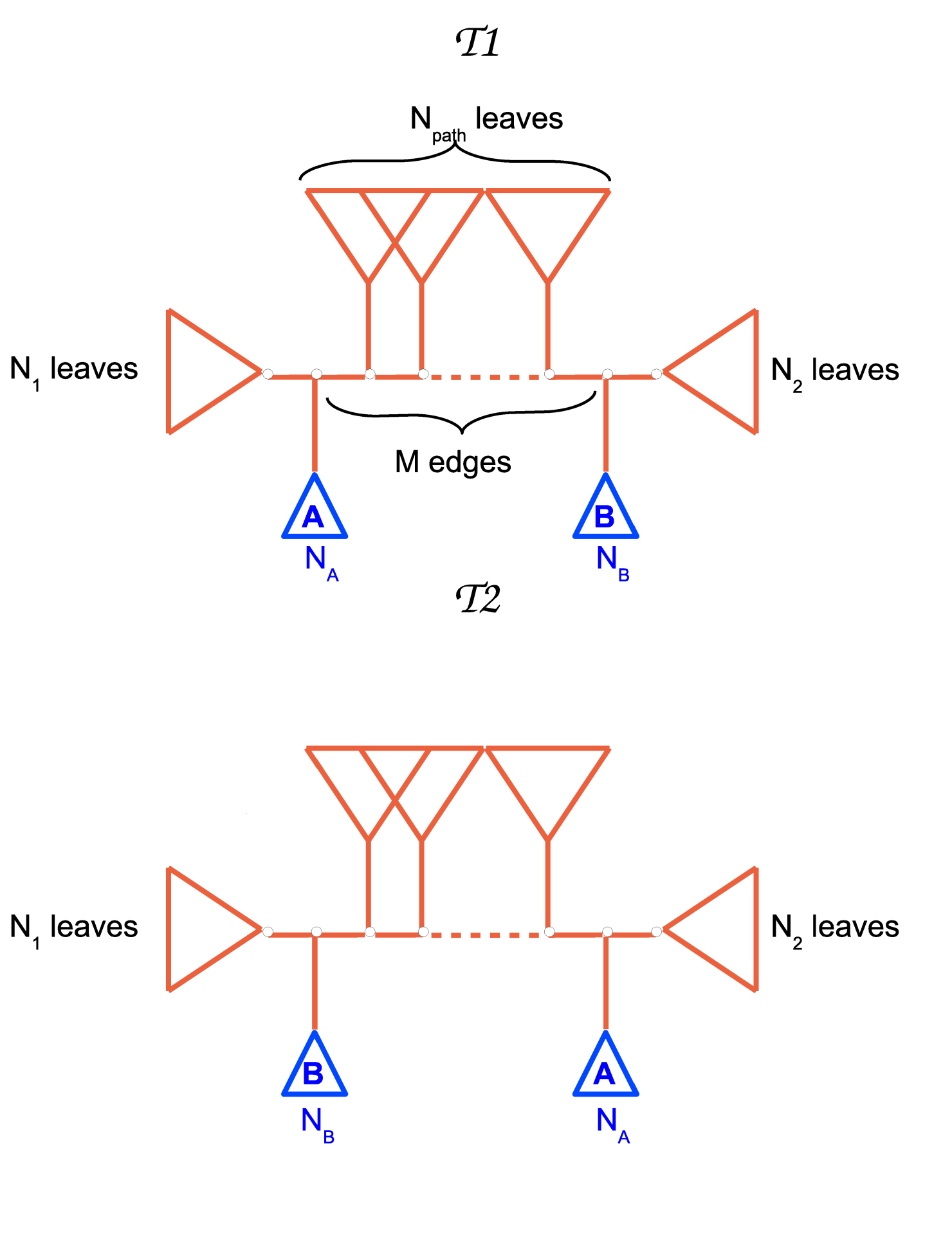} \end{center} \caption{{\bf Robinson-Foulds and Quartet Distance: errors due to a displacement of a couple of subtrees.} The trees ${\cal{T}}_{1}$ and  ${\cal{T}}_{2}$ are different because of the swap  of the subtrees {\bf A} and {\bf B}. While computing the distance between ${\cal{T}}_{1}$ and  ${\cal{T}}_{2}$, the {\bf Robinson-Foulds} distance detects all the $M$ edges in the path as errors, regardless of the size of the subtrees attached to them. The number of wrong butterflies quartets counted as errors with the {\bf Quartet Distance} is expressed by $N_1N_A(N_{path}N_B+N_{path}N_2+N_BN_2)+N_2N_B(N_1N_{path}+N_{path})N_A$: the QD thus depends on the size of the subtrees.} \label{QD_RF_displacement} \end{figure}

In~\cite{christensen_2005,Thesis_quartet_RF} a deep analysis of both RF and QD is reported, pointing out the different information the two measures convey.  In limiting cases, pairs of trees that have the same RF distance but very different QD, and vice-versa, are also shown.  In Fig.~\ref{QD_RF_displacement}, quoting an enlightening example in~\cite{christensen_2005,Thesis_quartet_RF}, we show how the RF and the QD measures weigh a swapping event of two subtrees in a tree. In this case the RF distance is equal to the number of edges in the path between the swapped subtrees, while the QD is sensitive to the size of the subtrees. The RF is then a good measure if we are interested in measuring how far apart subtrees are moved in one tree with respect to another. When we are interested instead in the size of the displaced subtrees, the quartet distance is a more adequate measure.

The Ethnologue classification provides a coarse grained grouping of subsets of languages, often leading to trees that are not fully resolved, i.e., that are not binary. For that reason, it is important to correct the biases suffered by the RF and QD distances while comparing binary with non binary trees.

\begin{figure}[htp] \begin{center} \includegraphics[width=0.8\textwidth]{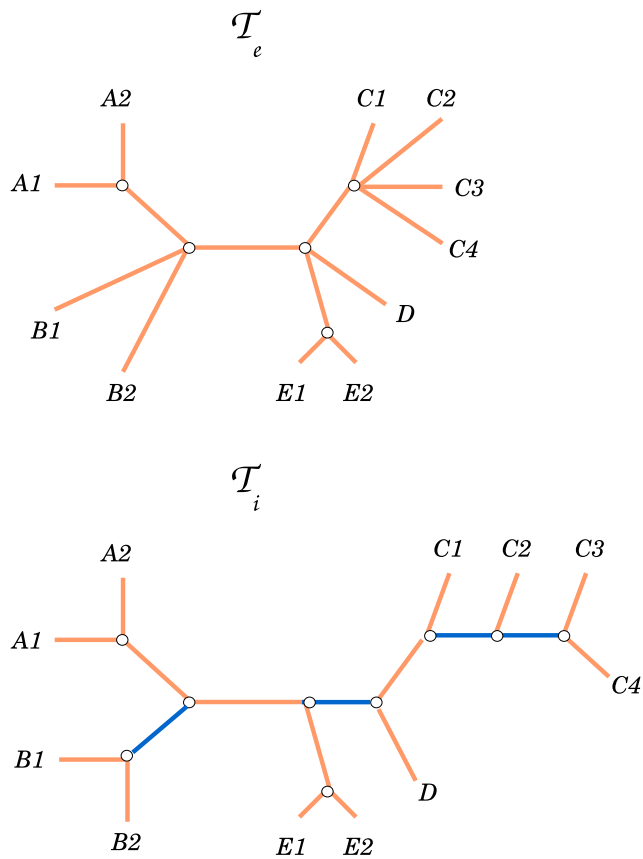} \end{center} \caption{{\bf Non-binary nodes: biases of errors. } The standard Robinson-Foulds distance and the Quartet Distance have a bias when comparing binary trees with non-binary classifications. The difference between tree ${\cal{T}}_{e}$ and ${{\cal{T}}_{i}}$ is that ${\cal{T}}_{i}$ shows a more fine grained classification. The two trees, however, are not conflicting, since ${\cal{T}}_{i}$ is simply a refinement of the classification ${\cal{T}}_{e}$.  The RF distance will count every internal edge (blue ones in ${{\cal{T}}_{i}}$) of this refinement as errors, since they are not in ${\cal{T}}_{e}$. The QD will count every quartet including the blue edges as errors, since all these quartets are stars in ${\cal{T}}_{e}$. The generalized measures we introduce correctly give a null score between ${\cal{T}}_{e}$ and ${{\cal{T}}_{i}}$ in the example.}  \label{cartoon_binary_non_binary} \end{figure}

Figure~\ref{cartoon_binary_non_binary} illustrates a situation when a binary tree (${{\cal{T}}_{i}}$) is compared with a non-binary one (${\cal{T}}_{e}$). Both the RF and the QD give a non zero distance between the two trees: some partitions of ${{\cal{T}}_{i}}$ are in fact not present in ${{\cal{T}}_{e}}$. It is important to consider, however, that in the case we are considering (algorithmic inference versus Ethnologue classification) non-binary classification is simply due to a lack of information or details that would lead to a finer classification. We would like to be able to distinguish intrinsic contradictions between reconstructed binary trees and the Ethnologue classifications from errors due to the low level of resolution of the Ethnologue trees.  It is with this aim in mind that we introduce a generalization of both the RF distance and the QD.

Let ${\cal{T}}_e$ be the Ethnologue (non necessarily binary) tree and ${\cal{T}}_i$ the inferred tree, then we define the Generalized Robinson-Foulds (GRF) score as:

\begin{equation} GRF({\cal{T}}_i,{\cal{T}}_e)=\frac{i({\cal{T}}_i)-e_{\mathrm{mod}}({\cal{T}}_i,{\cal{T}}_e)} {i({\cal{T}}_i)} \label{RF_gen} \end{equation}

\noindent where $i({\cal{T}}_i)$ denotes the number of internal edge of ${\cal{T}}_i$ and $e_{\mathrm{mod}}({\cal{T}}_i,{\cal{T}}_e)$  the number of bipartitions in ${\cal{T}}_i$  compatible with those in ${\cal{T}}_e$. Intuitively, a bipartition in ${\cal{T}}_i$ is said to be compatible with a bipartition in ${\cal{T}}_e$ if it does not contradict any of the bipartitions induced by cutting an edge in ${\cal{T}}_e$. More rigorously, the compatibility of a bipartition $b$ of ${\cal{T}}_i$ with the tree ${\cal{T}}_e$ is defined as follows. 
Let us call $b_1$ and $b_2$ the two sets defining $b$, and $a_1^i, a_2^i$ the two sets defining the $i$-th  bipartition of ${\cal{T}}_e$. The partition $b$ is compatible with the tree ${\cal{T}}_e$ if for each bipartition $i$ of ${\cal{T}}_e$, the following is true: $b_1 \subseteq a_1^i$, or $b_1 \subseteq a_2^i$, or $b_2 \subseteq a_1^i$, or $b_2 \subseteq a_2^i$. 

Let us note that the GRF is not symmetric in the two trees: this guarantees that a refinement edge is not counted as an error and the incomplete resolution of ${\cal{T}}_e$ does not affect the measure of the reliability of the reconstructed tree. We can verify that the GRF distance between ${\cal{T}}_i$ and ${\cal{T}}_e$  in figure~\ref{cartoon_binary_non_binary}is zero.

The QD is more straightforwardly generalized. We  define the Generalized Quartet Distance (GQD) score as:

\begin{equation} GQD({\cal{T}}_i,{\cal{T}}_e)=\frac{d({\cal{T}}_i,{\cal{T}}_e)} {\mathrm{norm}({\cal{T}}_e)} \label{QD_gen} \end{equation}

\noindent where $d({\cal{T}}i,{\cal{T}}e)$, as already introduced, denotes the number of different butterflies in ${\cal{T}}_i$ and ${\cal{T}}_e$. Again, this definition guarantees that all the star quartets in the Ethnologue trees will not be counted as errors. The normalization factor is equal to the number of butterfly quartets in ${\cal{T}}_e$: $\mathrm{norm}({\cal{T}}_e)=q({\cal{T}}_e)$, recalling the definition of $q({\cal{T}})$ given in eq.~\ref{QD}.

Let us stress again that both these generalized scores are neither symmetric or metric, since we are simply interested in quantifying the degree of accuracy of a binary tree with respect to an already known classification. With this definition, both the GQD and the GRF score give null scores if a classification tree is compared with one of its possible refinements, while one would get a score of $1$ for inferred trees in total disagreement with the classification. In the Supplementary Information we report a measure of the correlation of the accuracy of the trees reconstruction with the Ethnologue resolution, as measured both with the standard measures and with the generalized ones, showing how the last ones correctly remove the biases due to the incomplete Ethnologue classification.



\section*{Results}

\subsection*{Inferred trees vs. Ethnologue}

In this section we present the results of the comparison between the Ethnologue classification and the language trees inferred by state-of-the-art distance based algorithms. We first consider the ASJP database in order to perform a worldwide, i.e., large-scale, analysis.

Starting from the word lists of the ASJP project, we first estimated the distance matrices among all the languages in each family. We used both the LDN (\ref{LDN2}) and the LDND (\ref{LDND}) distances, so we had two classes of distance matrices as an input for distance-based algorithms.  We use three distance-based algorithms: {\em Neighbour-Joining (NJ)}~\cite{NJ}, {\em FastME}~\cite{fastme_2002} (belonging to the class of Balanced Minimum Evolution (BME) algorithms) and {\em FastSBiX}~\cite{sbix_2010,Fast-SBiX}, a recently introduced Stochastic Local Search algorithm. Each distance matrix was submitted as input to the three algorithms, which gives, for each language family, a total of six possible inferred trees.

To quantify the accuracy of the inferred trees, for each language family we computed the Generalized Robinson-Foulds score (GRF) and the Generalized Quartet Distance (GQD) of the inferred trees with the corresponding Ethnologue classifications.  Tables~\ref{GRF_mean_variance} and \ref{GQD_mean_variance} illustrate in an aggregate way the results obtained using the ASJP database. In particular we report, for each continent, the mean and the variance, across all the language families in that continent, of the values of the GRF and of the GQD between the inferred trees and the corresponding Ethnologue classifications, using both the LDN and the LDND distances. For each continent we considered all the language families present in the ASJP database.

As already mentioned, the GRF and the GQD are two complementary measures of the disagreement between the inferred tree and the expert classification. The GRF quantifies the percentage of wrong edges in the inferred trees, while the GQD counts how many quartets in the Ethnologue tree are different butterflies than in the reconstructed tree.  In both cases the performance of the different algorithms always look very similar, though in almost all cases the noise reduction made by FastSBiX corresponds to a slightly better ability in reconstructing the correct phylogenies. FastSBiX features indeed the lowest average scores and, in many cases, the lowest variances. As for the distance matrix, our results show how better performances are obtained, on average, by using the LDND distance (\ref{LDND}). The last column of the tables, named ``RANDOM'', shows the error one would have for a randomly reconstructed tree. This information is useful to correctly appreciate the algorithmic ability of inferring the correct phylogenetic relationships. While in fact we correct the distance measures in order to avoid biases due to non binary classification, it is evident that it is easier to be consistent with a very coarse grained classification than with a finer one. In order to take into account this observation, we can compare the errors made by the reconstruction algorithms with the errors a completely randomly constructed tree (with the same leaves) would feature. The RANDOM columns of tables~\ref{GRF_mean_variance} and \ref{GQD_mean_variance} report averages over 10 realizations of the GRF and the GQD between a randomly reconstructed tree and the Ethnologue classification.

\begin{figure}[htp] \begin{center} \includegraphics[width=\textwidth]{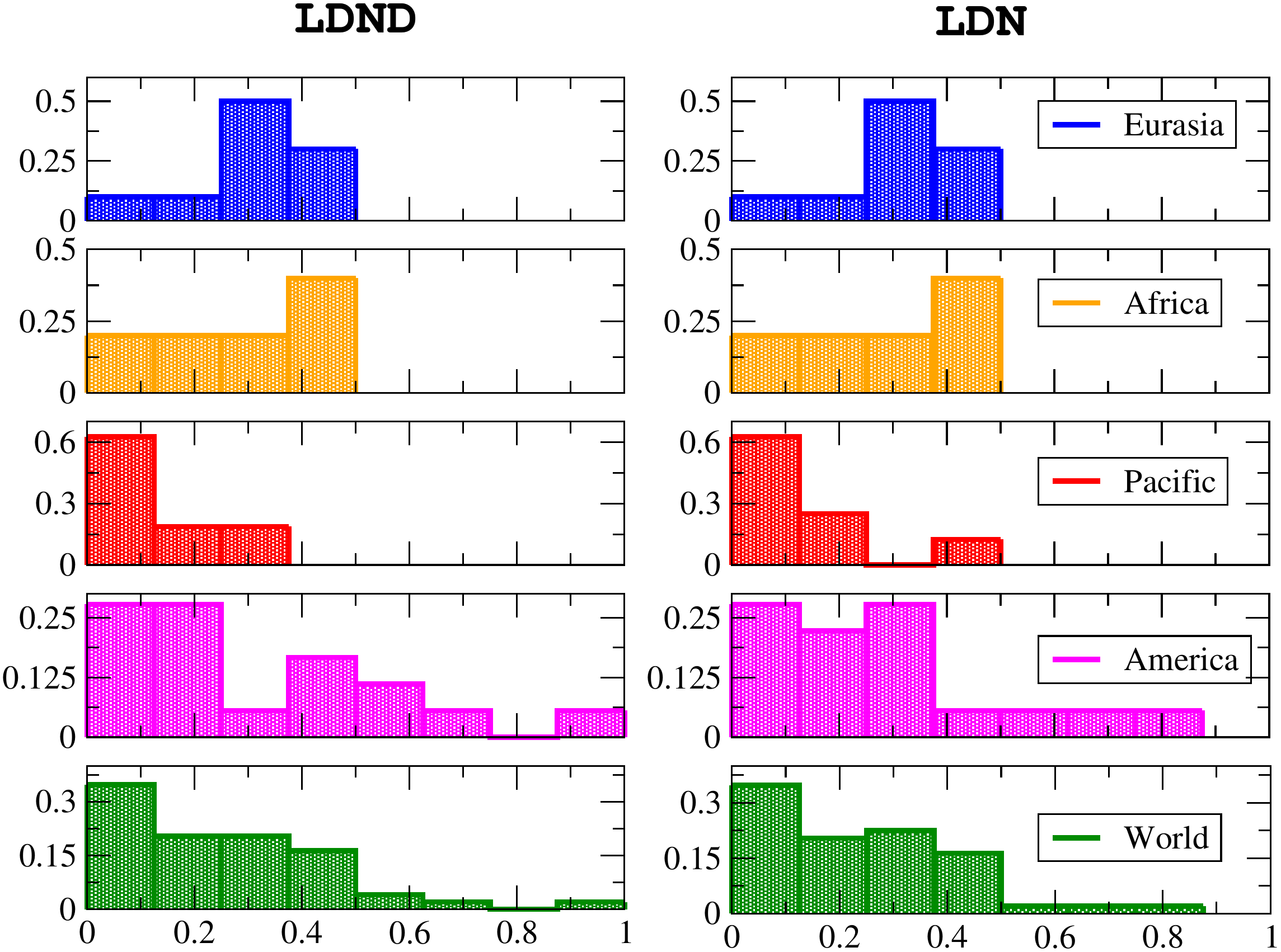} \end{center} \caption{ {\bf Accuracy histograms as measured with the Generalized Robinson-Foulds score (GRF).} For each continent and for the whole world we report the histograms of the GRF as measured over all the families spread on each specific region. We considered here only the FastSBiX algorithm that features slightly better performances with respect to the competing algorithms, and both the LDN (\ref{LDN2}) (right panel) and the LDND (\ref{LDND}) (left panel) definition of distance. The histograms are always peaked near zero, meaning that the rate of errors are always very low, but the variances are quite large.  These distributions do not discriminate the performances of the inference using LDN (\ref{LDN2}) or LDND (\ref{LDND}) definition of distances.}
\label{GRF_histo} \end{figure}

\begin{figure}[htp] \begin{center} \includegraphics[width=\textwidth]{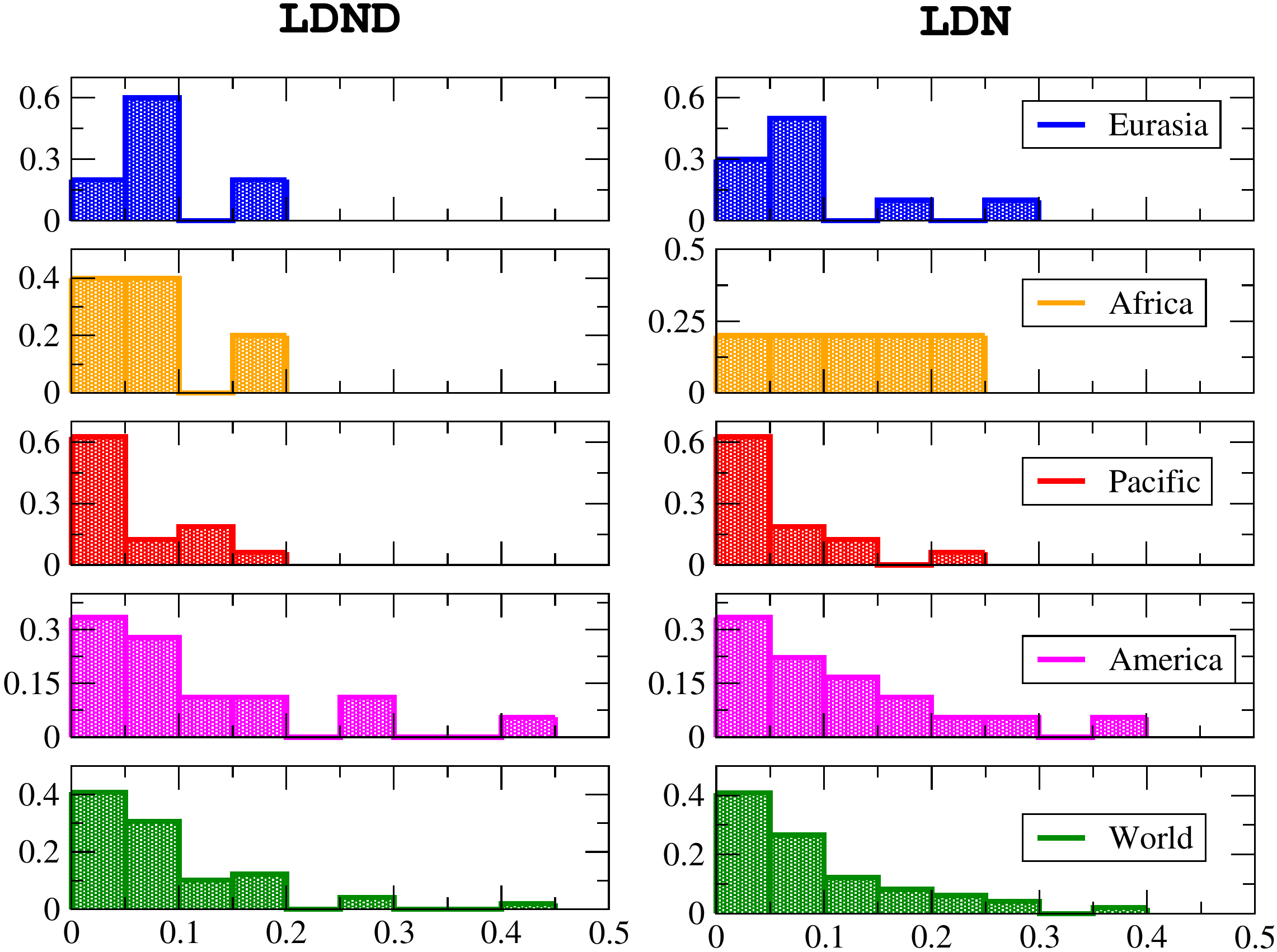} \end{center} \caption{ {\bf Accuracy histograms as measured with the Generalized Quartet Distance (GQD).} For each continent and for the whole world we report the histograms of the GQD as measured over all the families spread on each specific region. We considered here only the FastSBiX algorithm that features slightly better performances with respect to the competing algorithms, both with the LDN (\ref{LDN2}) (right panel) and the LDND (\ref{LDND}) (left panel) definition of distance. The histograms are always peaked near zero, meaning that the rate of errors are always very low. The distributions of the LDN-inferred trees, moreover, display larger variances than the LDND ones, this means that the latter definition allows for better performances in inferring languages trees with a distance-based approach. The overall variances are smaller with respect to the ones in fig.~\ref{GRF_histo}.} \label{GQD_histo} \end{figure}

Figures~\ref{GRF_histo} and \ref{GQD_histo} report the histograms of the accuracies obtained using the FastSBiX algorithm for each continent and worldwide: large fluctuations exist both within each continent and worldwide (The complete set of results for each language family and for all the accuracy scores is presented as Supporting Information in the tables~\ref{RF_table},~\ref{GRF_table},~\ref{QD_table} and~\ref{GQD_table}).

\begin{figure}[htp] \begin{center} \includegraphics[width=1.05\textwidth]{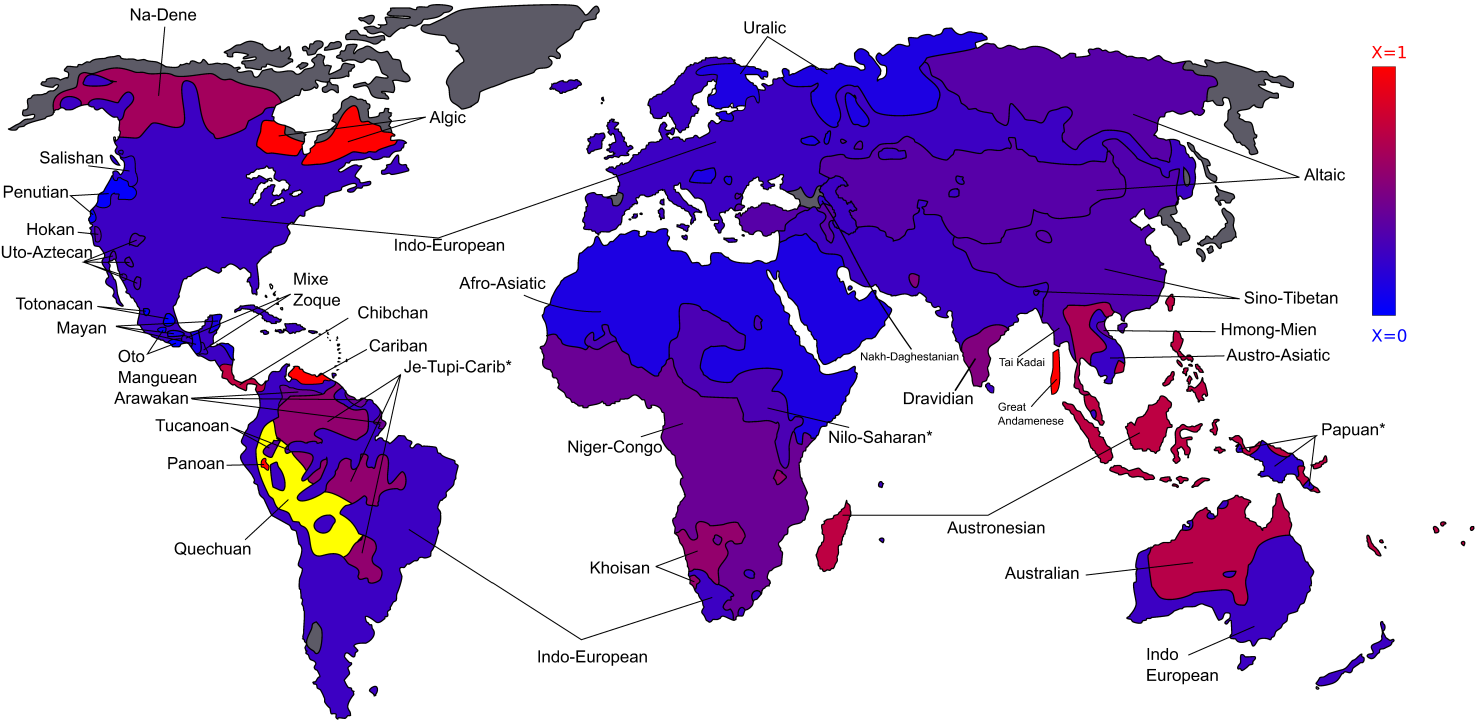} 

  \end{center} \caption{ {\bf Worldwide accuracy of the inferred language trees.} This map represents the level of accuracy of the FastSBiX algorithm on several language families throughout the world. The colors code the values of the Generalized Quartet Distance (GQD) between the trees inferred with the FastSBiX algorithm and the LDND definition of distance for each language family included in the ASJP database and the corresponding Ethnologue classifications. The GQD is normalized with the corresponding random value (see text for details). On the one hand blue regions corresponds to language families for which the inferred trees strongly agree with the Ethnologue classification. On the other hand red regions corresponds to poorly reconstructed language families. Yellow is for the families in which a random reconstruction would get a GQD score of zero, meaning that the Ethnologue classification has a null resolution (the corresponding tree is a star). Grey areas are those for which no data are present in the databases adopted for the reconstruction. Asterisks are for regions which include more than one family of languages. See the Supporting Information for the analogous maps obtained with different algorithms and different definitions of the distance between languages.}  \label{map_FastSBiX_LDND} \end{figure}

We finally give a pictorial view of the accuracy of the reconstruction algorithm across the planet. Figure~\ref{map_FastSBiX_LDND} illustrates the Generalized Quartet Distance for the different language families on the world map, normalized with the corresponding random value. More specifically, the color codes, for each family $f$, the following quantity:

\begin{equation}
X_f=2 \frac{GQD(f)}{GQD_{random}(f)}
\end{equation}

where $GQD_{random}(f)$ represents the mean value of the GQD obtained averaging over $10$ randomly reconstructed trees with the same leaves (languages) of the family $f$. $X_f$ quantifies the level of accuracy of the reconstruction with respect to a null model. The multiplicative factor $2$ is included for the sake of better visualization: $X_f=1$ indicates a $GQD(f)$ equal or higher to half of the random tree distance $GQD_{random}(f)$.

\begin{table}


\begin{center}\resizebox*{\textwidth}{!}{

\begin{tabular}{lccccccc}

\multicolumn{8}{c}{\bf GENERALIZED ROBINSON-FOULDS SCORE}\\\\ \hline

&\multicolumn{3}{|c|}{\bf LDN}&\multicolumn{3}{|c|}{\bf LDND}\\ \hline

&&\\

 &\bf Neighbour-Joining& \bf FastME& \bf FastSBiX& \bf Neighbour-Joining& \bf FastME& \bf FastSBiX& \bf RANDOM\\

&&\\

\hline

\multicolumn{8}{c}{ \bf AFRICA}\\ \hline

Mean& 0.2872& 0.2845& 0.2749& 0.2859& 0.2743& \bf{0.2729}& 0.7888\\

Variance& 0.0327& \bf{0.0322}& 0.0329& 0.0324& 0.0323& 0.0332& 0.1945\\

\hline

\multicolumn{8}{c}{ \bf EURASIA }\\ \hline

Mean& 0.3152& 0.3116& 0.2999& 0.3056& \bf{0.2930}& 0.2998& 0.9063\\

Variance& 0.0244& 0.0238& 0.0138& 0.0200& 0.0200& \bf{0.0108}& 0.0313\\

\hline

\multicolumn{8}{c}{ \bf PACIFIC }\\ \hline

Mean& 0.1228& 0.1271& 0.1092& 0.1200& 0.1178& \bf{0.1083}& 0.7282\\

Variance& \bf{0.0173}& 0.0182& 0.0181& 0.0174& 0.0177& 0.0177& 0.1422\\

\hline

\multicolumn{8}{c}{ \bf AMERICA }\\ \hline

Mean& 0.3084& 0.2885& \bf{0.2797}& 0.2972& 0.3080& 0.3023& 0.8949\\

Variance& 0.0673& 0.0600& \bf{0.0522}& 0.0673& 0.0726& 0.0654& 0.0525\\

\hline

\end{tabular}

}

\end{center}

 \caption{ {\bf Accuracy of the reconstructions as measured with the Generalized Robinson-Foulds (GRF).} For each continent we report the average and the variance of the GRF over all the languages spread on the continent. The different columns correspond to the two different ways of constructing the distance matrix (LDN and LDND) and to the three distance-based algorithms considered. The last column labelled RANDOM reports the results for the null model considered. See the main text for details.}

\label{GRF_mean_variance} \end{table}

\begin{table} 


\begin{center}\resizebox*{\textwidth}{!}{

\begin{tabular}{lccccccc}

\multicolumn{8}{c}{\bf GENERALIZED QUARTET DISTANCE}\\\\ \hline

&\multicolumn{3}{|c|}{\bf LDN}&\multicolumn{3}{|c|}{\bf LDND}\\ \hline

&&\\

 &\bf Neighbour-Joining& \bf FastME& \bf FastSBiX& \bf Neighbour-Joining& \bf FastME& \bf FastSBiX& \bf RANDOM\\

&&\\

\hline

\multicolumn{8}{c}{ \bf AFRICA }\\ \hline

Mean& 0.1379& 0.1872& 0.1379& 0.1094& 0.1048& \bf{0.0855}& 0.4781\\

Variance& 0.0072& 0.0164& 0.0069& 0.0047& 0.0045& \bf{0.0044}& 0.0601\\

\hline

\multicolumn{8}{c}{ \bf EURASIA }\\ \hline

Mean& 0.1911& 0.1787& 0.1721& 0.1716& 0.1676& \bf{0.1661}& 0.6437\\

Variance& 0.0378& 0.0387& 0.0399& 0.0386& 0.0385& 0.0355& \bf{0.0011}\\

\hline

\multicolumn{8}{c}{ \bf PACIFIC }\\ \hline

Mean& 0.0864& 0.0901& \bf{0.0662}& 0.0829& 0.0858& 0.0706& 0.4893\\

Variance& 0.0096& 0.0091& 0.0085& 0.0079& 0.0109& \bf{0.0070}& 0.0691\\

\hline

\multicolumn{8}{c}{ \bf AMERICA}\\ \hline

Mean& 0.1595& \bf{0.1536}& 0.1569& 0.1618& 0.1646& 0.1600& 0.6057\\

Variance& 0.0252& 0.0245& \bf{0.0235}& 0.0244& 0.0281& 0.0269& 0.0339\\

\hline

\end{tabular}

}

\end{center}

\caption{{\bf Accuracy of the reconstructions as measured with the Generalized Quartet Distance (GQD).} For each continent we report the average and the variance of the GQD over all the languages spread on the continent. The different columns correspond to the two different ways of constructing the distance matrix (LDN and LDND) and to the three distance-based algorithms considered. The last column labelled RANDOM reports the results for the null model considered. See the main text for details. }  \label{GQD_mean_variance} \end{table}

\subsection*{Effect of the database completeness and coverage}

In this section we consider how the length and the completeness of the lists of words affect the accuracy of the reconstruction. To this end, we restrict our analysis to the Austronesian family for which two different databases are available: the Automated Systematic Judgement Program (ASJP) and the Austronesian Basic Vocabulary Database (ABVD). The two databases mainly differ in two features: ASJP's lists include at most $100$ items for each language, while ABVD's lists includes up to $210$ words. In both cases, not all the languages in the family express all the meanings. As we have already pointed out in fig.~\ref{comparison}, while in the ASJP there are $40$ words shared by all the languages an additional $60$ words contained only in a small subset, in the ABVD database each word is shared at least by $50\%$ of the languages in the family.

In order to get a fair comparison, we isolate a subset of $305$ lists of words corresponding to languages shared by the two databases. The full list of languages is available in the Supporting Information, Table~\ref{ABVD_list}. These two classes of lists are used to infer phylogenetic trees of the corresponding languages to be compared with the Ethnologue classifications. Since the results of the previous section did not show a significant difference between the two definitions of distance matrix, here we only use the {\em LDN} distance which allows for faster computations. Further, we only consider the {\em FastSBiX} algorithm to reconstruct phylogenies, being the one that features slightly better performances, as shown in the previous section.

We start by investigating the effect of the length of the word-lists on the accuracy of the inference of evolutionary relationships among languages. To this end, for each of the two databases, we proceed as follows: for each meaning $i$ we compute the fraction $f_i$ of languages which contains a word for $i$. We sort these values in a decreasing order, obtaining a ranked list of words. We then consider different word-lists, obtained in the following way: we start with the $10$ most frequent words and we progressively add a constant number of words following the ranked list.

\begin{figure}[htp] \begin{center} \includegraphics[width=\textwidth]{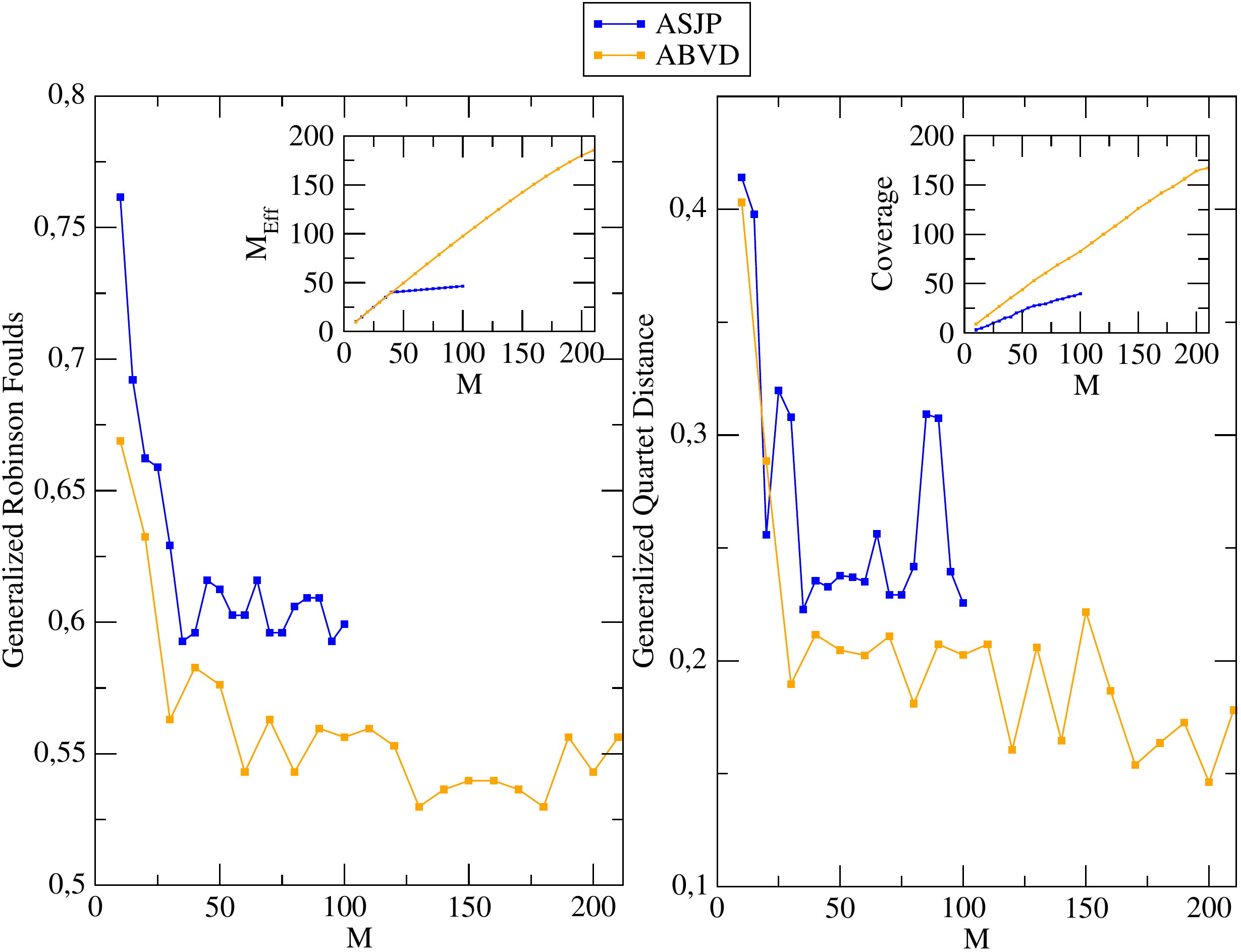} \end{center} \caption{{\bf Role of the word-list completeness and coverage.} {\bf (left)} the Generalized Robinson-Foulds (GRF) score between the inferred trees and the corresponding Ethnologue classification for the Austronesian family, vs. the number $M$ of most shared words, both for the ASJP and the ABVD databases. The inset reports the behaviour of $M_{eff}$, the effective number of most shared words, defines as follows. For each list $M_{eff}$ is the sum of all the value of $f_i$ for all the meanings in the list. In this way $M_{eff}$ quantifies the effective number of most shared meanings.  There is a strong correlation between $M$ and $M_{eff}$ for $M<40$. For $M>40$ $M_{eff}$ does not increase anymore in the ASJP database. This explains why the GRF does not decrease for $M>40$ for the ASJP database.  {\bf (right)} the Generalized Quartet Distance (GQD) between the inferred trees and the corresponding Ethnologue classification for the Austronesian family, vs. the number $M$ of most shared words, both for the ASJP and the ABVD databases. The inset reports the behaviour of the Coverage, which measures the degree of alignment of the word-lists for the different languages considered, vs. $M$ (see text for details about the definition of Coverage). Again there is a strong correlation between the Coverage and $M$. The distance-based algorithm used is FastSBiX with the LDN definition of distance.}  \label{completeness} \end{figure}

We compute the dissimilarity matrices by making use of only the reduced lists constructed as above, and we use those matrices as starting point for the reconstruction algorithm (we use the FastSBiX algorithm for all the results discussed below).  Fig.~\ref{completeness} reports the Generalized Robinson-Foulds score (left) and the Generalized Quartet Distance (right) between the inferred trees and the corresponding Ethnologue classifications, as a function of the number $M$ of chosen words, for both the AJSP and the ABVD databases. As a general trend, the number of errors decreases when the size of the word-lists considered increases. Though the large improvement of the accuracy occurs by adding the first $40$ or $50$ words, a slow improvement of the accuracy is always there if one keeps increasing the word-list size. This already points in the direction that, in order to improve the accuracy of the phylogenetic reconstruction, one has to increase the size of the word-lists. The accuracy obtained with the ABVD and ASJP databases are very similar when considering the first $M=40$ most shared words. Upon increasing $M$, ASJP does not feature any improvement while ABVD keeps improving its accuracy, although very slowly, when $M>40$. A possible explanation for this could be related to the presence, in the ASJP database, of meanings with a very low level of sharing (see inset of the left panel of Fig.~\ref{completeness} as well as Fig.~\ref{comparison}).

The value of $M_{eff}$ (see inset of the left panel of Fig.~\ref{completeness}) takes into account in how many languages a given meaning is expressed through a word. The missing information concerns whether pairs of languages have words for the same meaning.  Suppose two languages have words for the same number of meanings. This does not mean that the meaning expressed by words in each language are the same. If paradoxically the sets of meanings covered by the two languages had a null overlap, we wouldn't have data to construct distance matrices. It is thus interesting to measure the degree of overlap between the list of words of pairs of languages. To this end, we define each language $i$ as a binary vector $\vec{l_i}$ whose generic entry $l_i^a$ is $1$ if a word exists in that language for the meaning $a$ and $0$ otherwise. The overlap of two languages $l_i$ and $l_j$ is thus given by $\sum_a l_i^a l_j^a$. We define as level of coverage for a database the average overlap between all  pairs of languages:

\begin{equation}
\mathrm{Coverage} = \frac{2}{N (N-1)} \sum_{i \neq j} \sum_a l_i^a l_j^a,
\label{eq:coverage}
\end{equation}

\noindent where $N$ is the total number of languages considered, the index $a$ runs over all the meanings while the indices $i$ and $j$ run over the different languages. In this way the maximal value of the coverage is given by the total number of meanings $M$ we are considering.  The inset of the right panel of Figure~\ref{completeness} reports the curves for the Coverage as a function of $M$.  It is evident a strong correlation between $M$ and the Coverage both in the ASJP and ABVD databases. Notice that the maximal observed values of the coverage are well below the theoretical maximum ($100$) in the ASJP database and below the maximum ($210$) in the ABVD database.

The above results can be summarized by saying that the accuracy of the reconstructions strongly depends on the completeness (quantified by $M_{eff}$) as well as on the level of Coverage of the database considered. In the ASJP and ABVD databases $M$, $M_{eff}$ and the Coverage are strongly correlated and one observes a first substantial improvement of the accuracy for $M<40$ and a continuous, though slower, improvement for $M>40$ in the ABVD database, where $M_{eff}$ and the Coverage keeps increasing with $M$.

\section*{Discussion}

In this work we presented a quantitative investigation of the accuracy of distance-based methods in recovering evolutionary relations between languages. The quantification of the accuracy rests upon the computation of suitable distances between the inferred trees and the classifications made by experts (in our case the Ethnologue).

We introduced two generalized scores, the Generalized Robinson-Foulds score (GRF) and the Generalized Quartet Distance (GQD), which successfully allow for the comparison of binary trees and expert classifications. The generalizations were made necessary in order to take into account the biases due to the presence of non-binary nodes in the Ethnologue classifications, which came from a non fine-grained groupings of the languages. Our scores do not count every refinement as an error, while properly take in account every displacement of a language or wrong groupings with respect to the classifications. These scores are generalizations of standard measures; on the one hand the RF, which is a good measure if we are interested in measuring how far displaced pairs of subtrees have been moved around in one tree compared to another; on the other hand the QD is a more adequate measure whenever it is important to quantify the size of displaced subtrees. Our generalized scores inherit all these properties. Moreover, while in the GRF the stress is on the inferred trees, counting the percentage of wrong bipartitions in the reconstructed tree, in the GQD the stress is on the classification, since we are computing the percentage of correctly inferred quartets in the reconstructed tree.

Once properly defined the tools for the comparison, we conducted a thorough evalution of the accuracy of distance based methods on all the language families listed in the ASJP database. The analysis was carried out by adopting state-of-the art distance-based algorithms as well as two different definitions of distance between lists of words, the LDN (\ref{LDN2}) and the LDND (\ref{LDND}). In all the cases we obtained very robust results, which enabled us to draw some general conclusions. The two different definitions of distances between word-lists, LDN and LDND, almost guarantee the same accuracy for the inference of the trees of languages, as shown in tables~\ref{RF_table},~\ref{GRF_table},~\ref{QD_table},~\ref{GQD_table}, with the LDND definition allowing for a slightly better accuracy. The LDN, on the other hand, because of its lower computational complexity, allows for faster computations without a considerable loss of accuracy.  The length of the lists used to compute the distances between the languages strongly affects the accuracy of the reconstruction. The comparison between the two databases for the Austronesian family, the ASJP~\cite{ASJP} and the ABVD~\cite{ABVD} provides very important hints. The accuracy of the reconstruction always worsens if words with a low level of sharing are included; from this perspective it is always better to restrict the analysis to the meanings with an high Coverage instead of using all of them.

Fig.~\ref{map_FastSBiX_LDND} summarizes the accuracy of distance-based reconstruction algorithms for the different language families on the world map.  It is evident how at present the accuracy is satisfactory though highly heterogeneous across the different language families. Once the obvious bias is removed due to the finite Ethnologue resolution power, this heterogeneity has to be presumably ascribed to a non homogeneous level of completeness and coverage of the word-lists for specific language families.

In conclusion we provided the first extensive account of the accuracy of distance-based phylogenetic algorithms applied to the recontruction of worldwide language trees. The overall analysis shows as the effort devoted so far to the compilation of large-scale linguistic databases~\cite{ASJP,ABVD} already allows for very good reconstructions. We hope our survey could be an important starting point for further progress in the field, especially for language families for which the available databases are still incomplete or the corresponding Ethnologue classification still poorly resolved.




\section*{Acknowledgments}

The authors wish to warmly thank S$\varnothing$eren Wichmann for having provided support for the use of the ASJP database as well as for very interesting discussions. At the same time the authors wish to thank Simon J. Greenhill for having granted the permission of using the ABVD database.



\bibliography{phylo}


\newpage

\section{Supporting Information}

\subsection{Analysis of ASJP database and trees inference}

\begin{figure}[htp] \begin{center} \includegraphics[width=\textwidth]{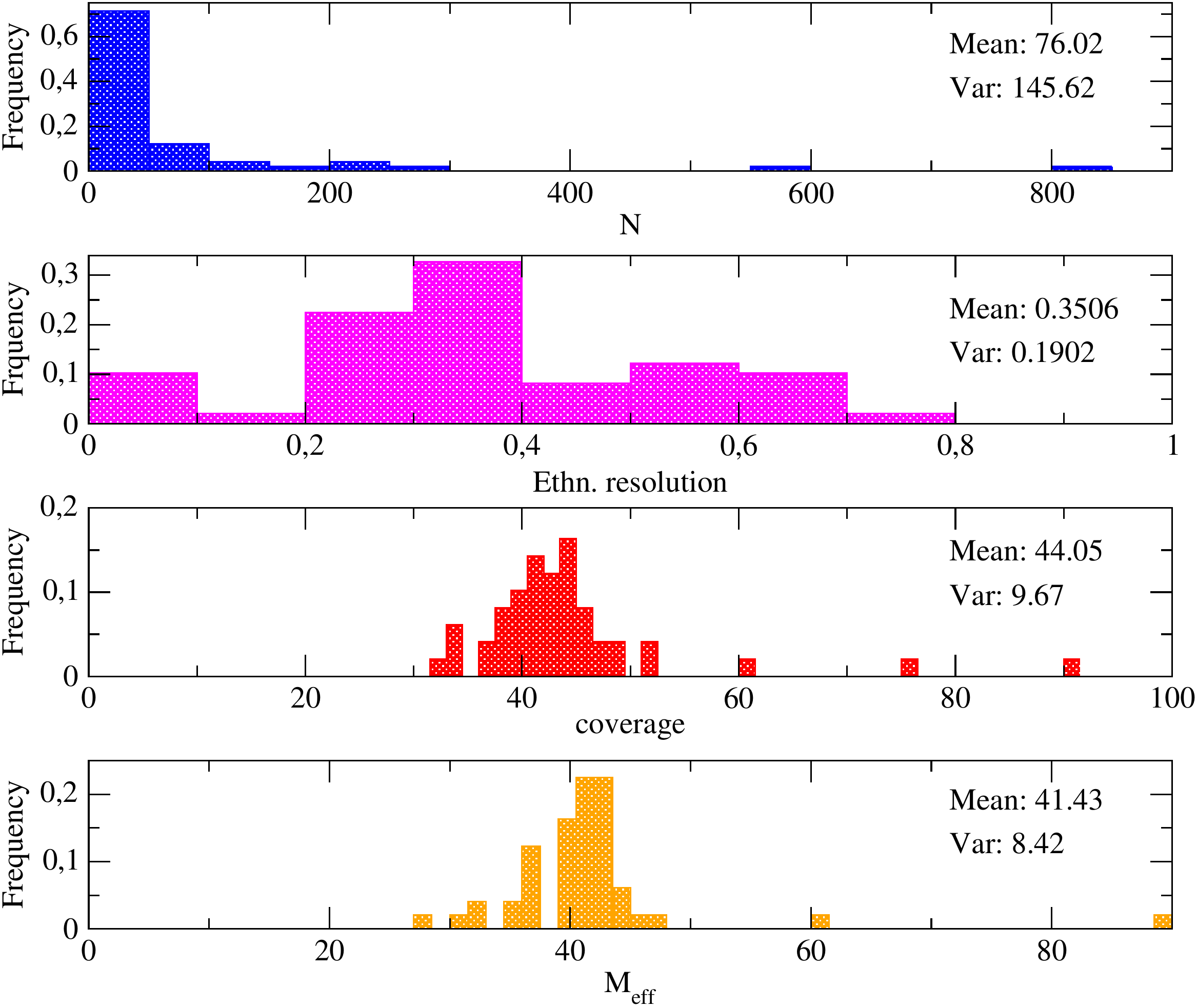} \end{center} \caption{ {\bf Histograms of ASJP and Ethnologue features.} Histograms of the number $N$ of languages in each family, the Ethnologue resolution for each family, $M_{eff}$ and the Coverage of the words lists in each family are reported.  While both $N$ and the Ethnologue resolution feature a great variability across the world family, $M_{eff}$ and the Coverage are almost constant in the lists of words related to language families in the ASJP database.}  \label{histos} \end{figure}

Here we discuss in details some quantitative features of the ASJP and Ethnologue databases.  Table~\ref{stat_data_families} summarizes, for each language family in the ASJP database, the number $N$ of languages, the corresponding resolution (as defined below) of the Ethnologue classification, and two properties of words lists discussed in the main text: $M_{eff}$ and the Coverage. Histograms showing the distribution across all the language families of the same quantities are shown in figure~\ref{histos}.  We note that, while $M_{eff}$ and Coverage are nearly constant for all the language families in the ASJP database, the number $N$ of languages and the Ethnologue resolution feature a great variability.  We quantify the resolution of the Ethnologue classification as $\frac{I-1}{N-2}$, where $I$ is the number of internal nodes in the classification tree and N is the number of leaves. With this definition, completely unresolved classifications, i.e., star trees, will result in a null resolution, while the resolution is equal to one for complete binary classifications. These are the values shown with a color code in Fig.~\ref{ethnologue_binary} in the main text.

For the sake of completeness, we recall the definition of $M_{eff}$ and Coverage. The former is defined as follows: we name $f_{meaning}$ the fraction of languages in each family which contain a word for $meaning$; $M_{eff}$ is simply the sum of $f_{meaning}$ over all the meanings expressed by a word in at least a language in the considered family.  The Coverage is a quantitative measure of the degree of overlap between the lists of words of pairs of languages, defined as:

\begin{displaymath}
\mathrm{Coverage} = \frac{2}{N (N-1)} \sum_{i \neq j} \sum_a l_i^a l_j^a,
\end{displaymath}

\noindent where we define each language $i$ as a binary vector $\vec{l_i}$, its generic entry $l_i^a$ being $1$ if a word exists in that language for the meaning $a$ and $0$ otherwise, and the sum is over all pairs of languages in the considered family.

We are now interested in analysing how these quantities affect the phylogeny reconstruction accuracy. To this end, we need to consider how the different measures of misclassification are in turn affected by the characteristics of the databases.

We then consider the accuracy of the inferred trees as measured respectively by the Robinson-Foulds distance, the Quartet Distance, the generalized Robinson-Foulds score and the generalized Quartet Distance. In particular, we investigate how the accuracy of the inferred trees is affected by the quantities considered above, namely the number $N$ of languages in the considered family, the resolution of the corresponding Ethnologue classification, the $M_{eff}$ and the Coverage.  In tables~\ref{pearson_1} and~\ref{pearson_2} we show the Pearson correlation coefficient (also known as {\rm Pearson's r}) between the distance of the inferred trees from the Ethnologue classification, as computed with the different criteria we proposed, and the quantities discussed above. In particular, table~\ref{pearson_1} shows results obtained considering the whole database while in table~\ref{pearson_2} we report results obtained by removing from the database those families with null Ethnologue resolution, i.e., for which the Ethnologue tree is a star.

In both cases, we observe a substantial difference between the standard RF and QD measures and their generalizations GRF and GQD.  The standard Robinson-Foulds distance features a positive Pearson correlation with the number $N$ of languages in a family, and both the standard Robinson-Foulds distance and Quartet Distance feature a strong negative Pearson correlation with the Ethnologue resolution. Both the GRF and GQD feature a Pearson coefficient with the Ethnologue resolution well below the significance threshold, correcting the biases in the misclassification measure due to lack of information in the Ethnologue database (see main text).

The reconstruction accuracy does not present correlations with $M_{eff}$ and the Coverage, the Pearson coefficient being below the significance threshold for the whole set of measures considered.  However, it is important to note that this lack of correlations is actually due to the homogeneity of the ASJP data set with respect to $M_{eff}$ and Coverage.  The histograms shown in Fig.~\ref{histos} for both $M_{eff}$ and Coverage are very peaked, with small variance: this absence of variability does not allow for the detection of correlations between such parameters and the accuracy of the reconstruction. In order to overcome this limitation we performed a comparative analysis of the ASJP and the ABVD databases for the Austronesian family (presented in the main text). The usage of a new database (the ABVD database) allows for the examination of words lists with very different values for both $M_{eff}$ and Coverage, revealing a strong dependence of the accuracy of the phylogenetic inference on such parameters.

In tables~\ref{RF_table}, \ref{GRF_table}, \ref{QD_table} and~\ref{GQD_table} extensive results for the accuracy of the inferred trees, as measured respectively by the Robinson-Foulds distance, the Generalized Robinson-Foulds score, the Quartet Distance and the Generalized Quartet Distance are reported.  The Robinson-Foulds distance, as already stressed, is sensitive to the length of the path between two displaced subtrees. Table~\ref{RF_table} shows how the reconstruction accuracy as measured by the RF does not depend on the definition of distance matrix (LDN vs. LDND) neither on the specific distance-based algorithm adopted.  As stressed in the main text, the use of the standard Robinson-Foulds distance (\ref{RF}) can lead to a systematic larger disagreement due to the presence of non binary internal nodes in Ethnologue trees, i.e., the existence of non fully resolved subgroups of a language family.  The generalized Robinson-Foulds score (\ref{RF_gen}) is not affected by this bias. The results obtained with the generalized RF score are shown in table~\ref{GRF_table}.  Next we consider the accuracy of the inferred trees as measured by the Quartet Distance. If we take into account the standard definition of the QD (table~\ref{QD_table}), the accuracy of the inferred trees result quite low, the average distance between the inferred and expert classification being around $45\%$. The adoption of the generalized Quartet Distance score (\ref{GQD_table}) allows to remove the biases due to the presence of star quartets in the classification trees. The generalized QD scores are reported in table~\ref{GQD_table}. The accuracy of the inferred trees in this case turns out to be much higher, with an average fraction of disagreement lower than $10\%$ and with large fluctuations from a minimum of zero to a maximum of roughly $30\%$ for the Panoan family.

\subsection{Analysis of ASJP and ABDV databases for the Austronesian Family}

We give here some supplementary information about the analysis on the Austronesian Family we presented in the main text. The set of 305 Austronesian languages taken in account is presented in Table~\ref{ABVD_list}, the name of languages are the ones shown in \url{http://language.psy.auckland.ac.nz/austronesian/}.

The Coverage of the ABVD lists for this set is of $167.32$, for the ASJP is $39.83$. The value of the $M_{eff}$ is $185,23$ for the ABVD database and $46,49$ for the ASJP. The ABVD thus features higher values for both parameters.

In Table~\ref{ABVD_accuracy} we show the distance from the Ethnologue classification, as computed by all the four measures adopted (Robinson-Foulds, Generalized Robinson-Foulds, Quartet-Distance, Generalized Quartet-Distance), of the most accurate tree inferred by using the ABVD list and the most accurate tree inferred by using the ASJP list. The most accurate tree is intended to be the tree, within the ones inferred by the three considered algorithms, that features lower distance from the Ethnologue classification.  The LDN definition of distance between languages, which allows for faster computations, has been used here.

All the four tree distances used, point that the inference made by using ABVD lists is more accurate than the one made by the use of ASJP lists: this is a consequence of the higher values of the Coverage and of $M_{eff}$ featured by the ABVD database (see main text). FastSBiX appears again to be the best distance-based algorithm for tree reconstruction, both considering the GRF and GQD distances.

\subsection{World maps}

\begin{figure}[htp] \begin{center} 

\includegraphics[width=\textwidth]{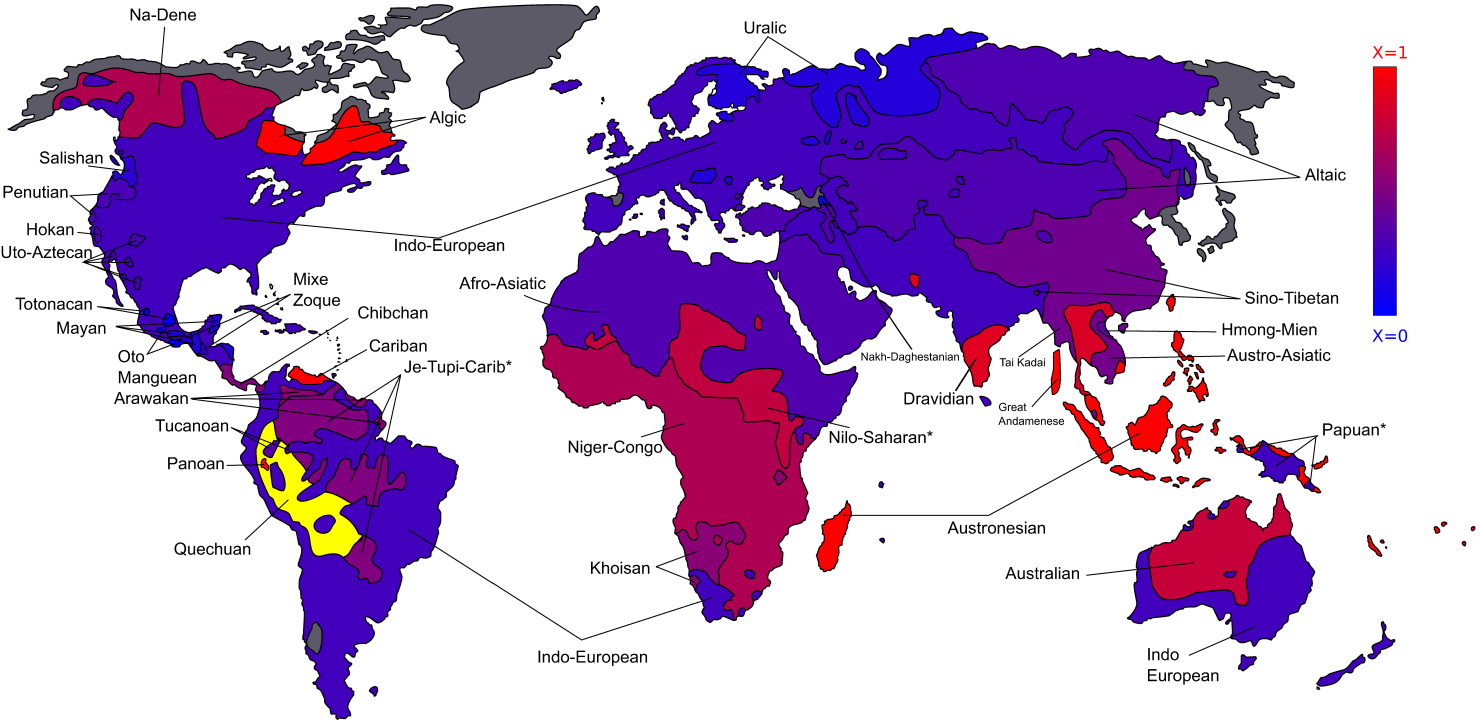}

\end{center} 

\caption{ {\bf Accuracy of the reconstruction across the planet. NJ, LDN matrix.} This map represents the level of accuracy of the Neighbour-Joining algorithm on several language families throughout the world. Trees Inferred with LDN matrices. The legend is the same of Fig.\ref{map_FastSBiX_LDND}.}

\label{map_NJ_LDN}

\end{figure}

\begin{figure}[htp] \begin{center} 

\includegraphics[width=\textwidth]{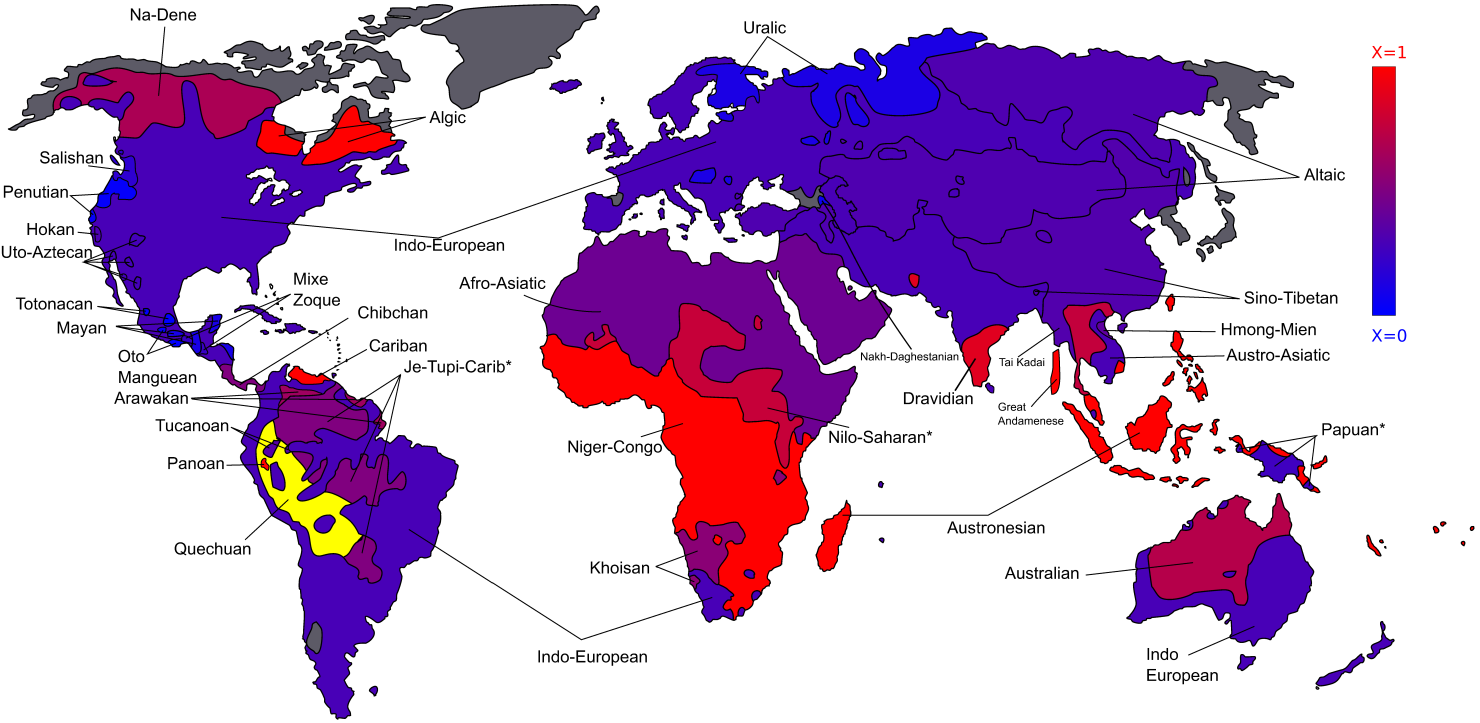}

\end{center} 

\caption{ {\bf Accuracy of the reconstruction across the planet. FastME, LDN matrix.} This map represents the level of accuracy of the FastME algorithm on several language families throughout the world. Trees Inferred with LDN matrices. The legend is the same of Fig.\ref{map_FastSBiX_LDND}.}

\label{map_FastME_LDN}

\end{figure}

\begin{figure}[htp] \begin{center} 

\includegraphics[width=\textwidth]{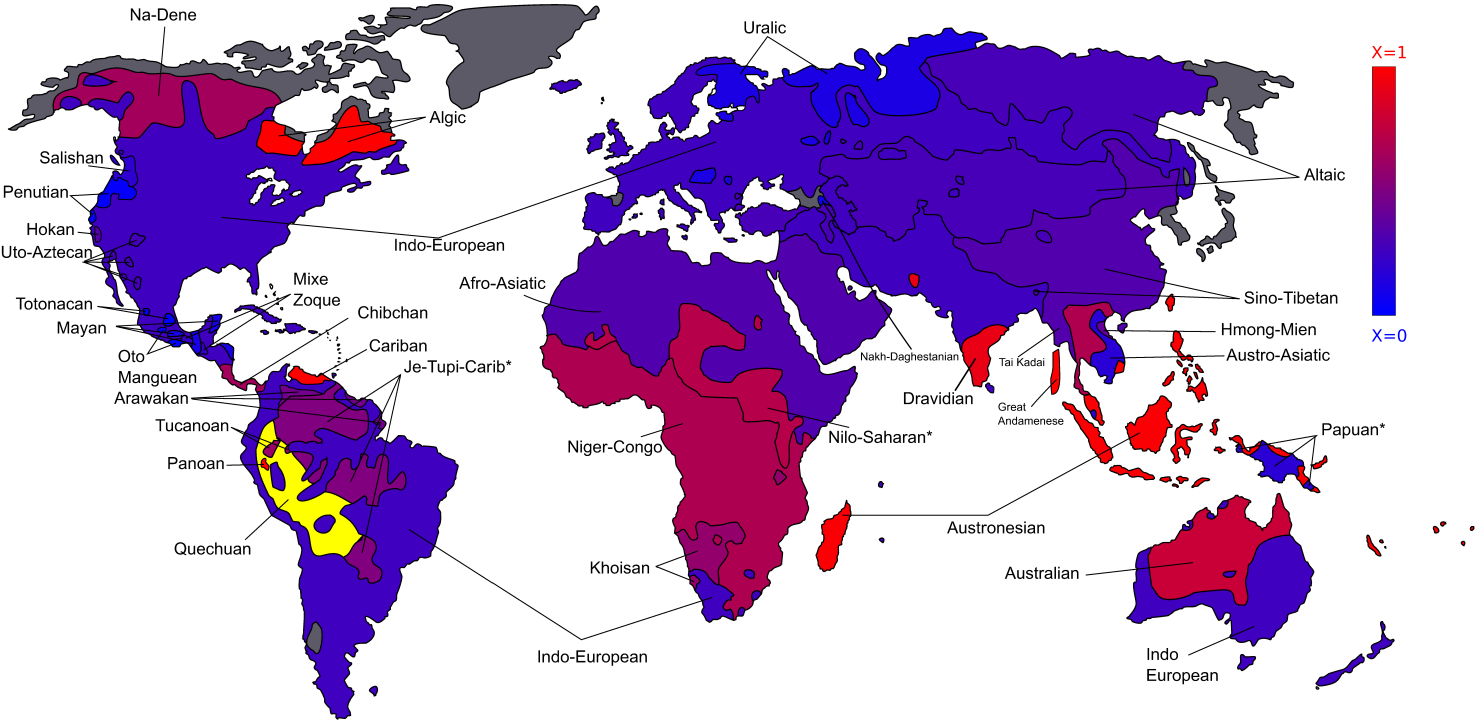}

\end{center} 

\caption{ {\bf Accuracy of the reconstruction across the planet. FastSBiX, LDN matrix.} This map represents the level of accuracy of the Fast-SBiX algorithm on several language families throughout the world. Trees Inferred with LDN matrices. The legend is the same of Fig.\ref{map_FastSBiX_LDND}.}

\label{map_FastSBiX_LDN}

\end{figure}

\begin{figure}[htp] \begin{center} 

\includegraphics[width=\textwidth]{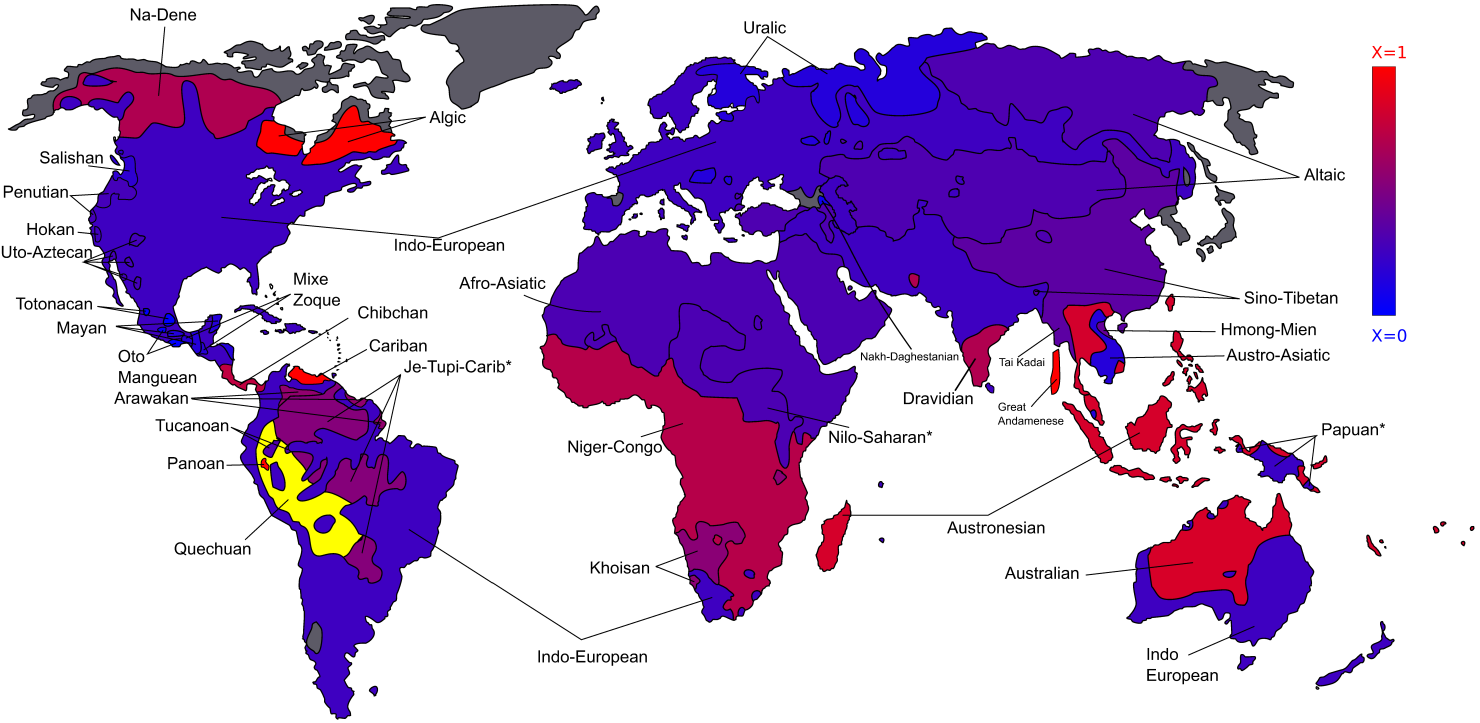}

\end{center} 

\caption{ {\bf Accuracy of the reconstruction across the planet. NJ, LDND matrix.} This map represents the level of accuracy of the Neighbour-Joining algorithm on several language families throughout the world. Trees Inferred with LDND matrices. The legend is the same of Fig.\ref{map_FastSBiX_LDND}.}

\label{map_NJ_LDND}

\end{figure}

We report here world maps showing the Generalized Quartet Distance between inferred trees and Ethnologue classifications. Trees inferred starting from LDN matrices are shown in Fig.~\ref{map_NJ_LDN} (with Neighbour-Joining), Fig.~\ref{map_FastME_LDN} (FastME) and in Fig.~\ref{map_FastSBiX_LDN}. Trees inferred starting from LDND matrices are shown in Fig.~\ref{map_NJ_LDND} (Neighbour-Joining), Fig.~\ref{map_FastME_LDN} (FastME). In the main text we have shown the map of the LDND-inferred trees with FastSBiX. We recall that the colors code, for each family $f$, the following quantity $X_f=2*GQD(f)/GQD_{random}(f)$, where $GQD_{random}(f)$ represents the value of the GQD obtained as average over $10$ random trees with the same number of leaves (languages) of the family $f$ (see main text).  $X_f$ quantifies the level of accuracy of the reconstruction with respect to a null model. Blue families are those for which the accuracy is very high while for red families the accuracy is very low, i.e., smaller than half the the random value. Yellow regions on the maps are related to (non-significant) families with null Ethnologue resolution, for which a random reconstruction would get a null value of the GQD.

The difference of the accuracy of the different algorithms is more evident in some regions such as the whole Africa, the Oceania and the east-Europe, whereas areas such as the whole America do not exhibit big differences in all the maps. This visual analysis immediately reveals the main conclusions we have drawn in the main text. Recalling that darker colors (i.e., color with a higher percentage of blue) point to a better accuracy, we see that the sensible regions always get darker while going from LDN maps to LDND ones. This behaviour reveals a slightly better accuracy achieved with the former definition of distance between lists of words. The effect of the different distance-based algorithms used for the reconstruction is visibly more evident. Neighbour-Joining maps displays more red regions than FastME maps; these sensible regions always get darker when observing FastSBiX maps. This visual analysis, thus, enlightens once again the suitability of the noise-reduction procedure used by FastSBiX to infer the correct topology of language trees.

We finally list all the regions where we included the average statistics of more than one language family: In the Nilo-Saharan region we included Kadugli and Nilo-Saharan; in the Papuan-labelled region we included Bosavi, Eleman, Kiwaian, Sko, Western Fly, Marind, Sepik, West Papuan, Trans-New Guinea, Torricelli, Morehead and Upper Maro Rivers, Lakes Plain, Border, Lower Sepik-Ramu.

\begin{table}[htp] 

\begin{center}\resizebox*{14cm}{!}{

\begin{tabular}{lcccc}

\hline

\bf Family& \bf N & \bf Ethn. Resolution& \bf Coverage& \bf $M_{Eff}$\\

\hline

Afro-Asiatic& 227& 0.38& 43.07& 39.66\\

Algic& 28& 0.22& 43.93& 40.34\\

Altaic& 75& 0.31& 42.47& 41.4\\

Arawakan& 48& 0.21& 43.73& 41.06\\

Australian& 186& 0.32& 40.29& 37.05\\

Austro-Asiatic& 52& 0.64& 60.27& 47.54\\

Austronesian& 833& 0.35& 43.93& 40.1\\

Border& 16& 0.14& 33.94& 32.66\\

Bosavi& 15& 0& 39.33& 39.72\\

Cariban& 19& 0.53& 44.16& 42.4\\

Chibchan& 20& 0.31& 47.67& 43.14\\

Dravidian& 21& 0.79& 45.05& 36.15\\

Eleman& 10& 0.63& 44.3& 41.49\\

Great Andamanese& 10& 0.25& 38.3& 39.91\\

Hmong-Mien& 14& 0.25& 41& 43.08\\

Hokan& 24& 0.57& 46.22& 42.29\\

Indo-European& 210& 0.36& 43.69& 41.16\\

Kadugli& 11& 0& 41& 44\\

Khoisan& 16& 0.64& 41& 42.67\\

Kiwaian& 15& 0& 39& 39.69\\

Lakes Plain& 26& 0.21& 37.19& 35.19\\

Lower Sepik-Ramu& 20& 0.5& 31.95& 28.21\\

Macro-Ge& 24& 0.32& 48.5& 42.72\\

Marind& 32& 0.33& 34.09& 30.79\\

Mayan& 75& 0.33& 75.03& 60.09\\

Mixe-Zoque& 14& 0.7& 91& 89.09\\

Morehead and Upper Maro Rivers& 17& 0.7& 34& 31.79\\

Na-Dene& 22& 0.6& 46.45& 42.42\\

Nakh-Daghestanian& 32& 0.43& 41& 41.29\\

Niger-Congo& 558& 0.4& 41.34& 39.89\\

Nilo-Saharan& 113& 0.54& 42.07& 40.38\\

Oto-Manguean& 60& 0.3& 43.02& 40.73\\

Panoan& 18& 0.31& 41& 42.35\\

Penutian& 21& 0.31& 52.43& 44.29\\

Quechuan& 18& 0.06& 44.33& 42.55\\

Salishan& 12& 0.4& 51& 45.45\\

Sepik& 26& 0.25& 37.81& 36.62\\

Sino-Tibetan& 141& 0.35& 42.7& 41.12\\

Sko& 14& 0.33& 39& 41.57\\

Tai-Kadai& 56& 0.31& 42.11& 40.78\\

Torricelli& 31& 0.21& 37.06& 35.63\\

Totonacan& 14& 0.25& 41& 44.44\\

Trans-New Guinea& 293& 0.31& 40.02& 36.01\\

Tucanoan& 14& 0.59& 47.32& 42.92\\

Tupian& 47& 0.24& 44.75& 41.09\\

Uralic& 23& 0.43& 48.74& 42.71\\

Uto-Aztecan& 81& 0.34& 45.5& 41.37\\

West Papuan& 34& 0.22& 38.21& 36.17\\

Western Fly& 39& 0& 37.77& 37.01\\

\hline

\end{tabular}

}

\end{center}

\caption{ {\bf Statistical Properties of the families in the ASJP database}. In this table we report the number $N$ of languages and the resolution of the Ethnologue classification for each family in the ASJP data set. We also show two statistical properties of the lists of words: the $M_{eff}$ of words and the Coverage. See the text for the definitions.}

\label{stat_data_families}

\end{table}

\begin{table}[htp]

\begin{center}{

\begin{tabular}{lcccc}

\hline

&\bf N & \bf Ethn. resolution& \bf Coverage& \bf $M_{eff}$\\

\hline

{\bf RF} - FastME \hspace{1cm}& 0.3514& -0.1298& -0.1542& -0.1903\\

{\bf RF} - NJ \hspace{1cm}& 0.3097& -0.1459& -0.1654& -0.2228\\

{\bf RF} - FastSBiX \hspace{1cm}& 0.3488& -0.2108& -0.1045& -0.1283\\

\hline

{\bf QD} - FastME \hspace{1cm}& -0.1458& -0.6704& -0.2605& -0.1066\\

{\bf QD} - NJ \hspace{1cm}& -0.1847& -0.6901& -0.2623& -0.1120\\

{\bf QD} - FastSBiX \hspace{1cm}& -0.211& -0.6717& -0.2494& -0.0926\\

\hline

{\bf GRF} - FastME \hspace{1cm}& 0.1925& 0.3414& 0.0453& 0.0024\\

{\bf GRF} - NJ \hspace{1cm}& 0.1829& 0.3950& 0.0833& 0.0512\\

{\bf GRF} - FastSBiX \hspace{1cm}& 0.1869& 0.3543& 0.1230& 0.0985\\

\hline

{\bf GQD} - FastME \hspace{1cm}& 0.0686& 0.1079& -0.0923& -0.0749\\

{\bf GQD} - NJ \hspace{1cm}& 0.1338& 0.1352& -0.0855& -0.0599\\

{\bf GQD} - FastSBiX \hspace{2cm}& 0.0084& 0.1392& -0.0597& -0.0256\\

\hline

\end{tabular}

}

\end{center}

\caption{ {\bf Pearson's r between reconstruction accuracy and parameters of ASJP and Ethnologue databases. } This table shows the Pearson's r between the accuracy (as measured by the RF, the QD, the GRF and the GDQ) of the  trees inferred with FastSBiX, FastME and Neighbour-Joining, starting from the LDND matrices, and different properties of the lists of words of each family in the ASJP database. In particular, we compute the Pearson's r between the estimated accuracy of the inferred trees and the number $N$ of languages in a family (first column), the Ethnologue Resolution (second column), the Coverage of words in lists (third column) and the effective length of the word-lists $M_{eff}$ (last column).}

\label{pearson_1}

\end{table}

\begin{table}[htp]

\begin{center}{

\begin{tabular}{lcccc}

\hline

 &\bf N & \bf Ethn. resolution& \bf Coverage& \bf $M_{eff}$\\

\hline

{\bf RF} - FastME & 0.3593& -0.1370& -0.1497& -0.1837\\

{\bf RF} - NJ & 0.3155& -0.1651& -0.1624& -0.2171\\

{\bf RF} - FastSBiX & 0.3604& -0.2213& 0.0959& -0.1203\\

\hline

{\bf QD} - FastME & -0.0715& -0.4126& -0.2459& -0.1231\\

{\bf QD} - NJ & -0.1309& -0.4476& -0.2499& -0.1322\\

{\bf QD} - FastSBiX & -0.1713& -0.4132& -0.1022& -0.1023\\

\hline

{\bf GRF} - FastME & 0.1542& 0.1403& -0.0078& -0.0120\\

{\bf GRF}  - NJ & 0.1445& 0.2216& 0.0355& 0.0415\\

{\bf GRF}  - FastSBiX & 0.1482& 0.1591& 0.0780& 0.0931\\

\hline

{\bf GQD} - FastME & 0.0285& -0.1111& -0.1433& -0.0907\\

{\bf GQD}  - NJ & 0.0994& -0.0636& -0.1334& -0.0740\\

{\bf GQD}  - FastSBiX & -0.0321& -0.0511& 0.0883& -0.0374\\

\hline

{\bf GQD/GQD$_{\mathrm{random}}$} - FastME& 0.1127& -0.1192& -0.1507& -0.0990\\

{\bf GQD/GQD$_{\mathrm{random}}$} - NJ& 0.1969& -0.0724& -0.1394& -0.0821\\

{\bf GQD/GQD$_{\mathrm{random}}$} - FastSBiX& 0.0348& -0.0576& 0.0746& -0.0445\\

\hline

 {\bf GRF/GRF$_{\mathrm{random}}$ }    - FastME & 0.1157& 0.0779& -0.0391& -0.0229\\

 {\bf GRF/GRF$_{\mathrm{random}}$  }   - NJ & 0.1084& 0.1610& 0.0026& 0.0276\\

 {\bf GRF/GRF$_{\mathrm{random}}$  }  - FastSBiX & 0.1097& 0.0994& 0.0434& 0.0782\\

 \hline

\end{tabular}

}

\end{center}

\caption{ {\bf Pearson's r between reconstruction accuracy and parameters of ASJP and Ethnologue databases. } We show here the same quantities as in table~\ref{pearson_1}, but the correlations are now computed without taking in account families in which the Ethnologue database does not provide a classification (it results in a star tree). Here, the measures $GQD/GQD_{\mathrm{random}}$ and  $GRF/GRF_{\mathrm{random}}$ are also taken into account. They allow for an analysis of the correlations of the accuracy of the inference with respect to a null model. Correlations are still very low for all the considered quantities, FastSBiX revealing for all of them the lowest correlation.}

\label{pearson_2}

\end{table}

\begin{table}[htp] 


\begin{center}\resizebox*{14cm}{!}{

\begin{tabular}{lccccccc}

\multicolumn{8}{c}{\bf ROBINSON-FOULDS DISTANCE}\\\\ \hline

&\multicolumn{3}{|c|}{\bf LDN}&\multicolumn{3}{|c|}{\bf LDND}\\ \hline

&&\\

 &\bf Neighbour-Joining& \bf FastME& \bf FastSBiX& \bf Neighbour-Joining& \bf FastME& \bf FastSBiX& \bf RANDOM\\

&&\\

\hline

\multicolumn{8}{c}{ \bf AFRICA }\\ \hline

Khoisan& \bf{0.4688}& \bf{0.4688}& \bf{0.4688}& \bf{0.4688}& \bf{0.4688}& \bf{0.4688}& 0.9905\\

Niger-Congo& 0.4964& 0.4946& \bf{0.4910}& 0.5000& 0.5018& 0.4964& 0.9995\\

Nilo-Saharan& 0.4292& 0.4292& 0.4027& 0.4204& 0.4027& \bf{0.3938}& 0.9952\\

Kadugli& \bf{0.3636}& \bf{0.3636}& \bf{0.3636}& \bf{0.3636}& \bf{0.3636}& \bf{0.3636}& 1.0000\\

Afro-Asiatic& 0.4626& 0.4537& 0.4537& 0.4626& \bf{0.4449}& \bf{0.4449}& 0.9994\\

\hline

\multicolumn{8}{c}{ \bf EURASIA }\\ \hline

Indo-European& 0.5500& \bf{0.5310}& 0.5405& 0.5500& 0.5357& 0.5405& 1.0000\\

Uralic& 0.3696& 0.3696& 0.3696& 0.3696& 0.3696& \bf{0.3261}& 0.9862\\

Altaic& \bf{0.5400}& \bf{0.5400}& \bf{0.5400}& \bf{0.5400}& \bf{0.5400}& 0.5533& 0.9979\\

Dravidian& 0.5000& 0.5000& \bf{0.3571}& 0.4524& 0.4524& \bf{0.3571}& 0.9879\\

Nakh-Daghestanian& \bf{0.2813}& \bf{0.2813}& \bf{0.2813}& \bf{0.2813}& \bf{0.2813}& \bf{0.2813}& 1.0000\\

Sino-Tibetan& 0.5390& 0.5248& \bf{0.5177}& 0.5390& \bf{0.5177}& \bf{0.5177}& 0.9989\\

Hmong-Mien& \bf{0.3571}& \bf{0.3571}& \bf{0.3571}& \bf{0.3571}& \bf{0.3571}& \bf{0.3571}& 1.0000\\

Tai-Kadai& \bf{0.4911}& \bf{0.4911}& \bf{0.4911}& \bf{0.4911}& \bf{0.4911}& \bf{0.4911}& 1.0000\\

Great Andamanese& \bf{0.4000}& \bf{0.4000}& \bf{0.4000}& \bf{0.4000}& \bf{0.4000}& \bf{0.4000}& 1.0000\\

Austro-Asiatic& 0.3462& 0.3462& 0.3654& \bf{0.3269}& 0.3462& 0.3654& 0.9975\\

\hline

\multicolumn{8}{c}{ \bf PACIFIC}\\ \hline

Austronesian& 0.5258& 0.5563& 0.5258& \bf{0.5246}& 0.5330& \bf{0.5246}& 0.9994\\

Border& 0.4375& 0.4375& \bf{0.3750}& 0.4375& 0.4375& \bf{0.3750}& 1.0000\\

Bosavi& \bf{0.4000}& \bf{0.4000}& \bf{0.4000}& \bf{0.4000}& \bf{0.4000}& \bf{0.4000}& 1.0000\\

Kiwaian& \bf{0.4000}& \bf{0.4000}& \bf{0.4000}& \bf{0.4000}& \bf{0.4000}& \bf{0.4000}& 1.0000\\

Eleman& \bf{0.1000}& \bf{0.1000}& \bf{0.1000}& \bf{0.1000}& \bf{0.1000}& \bf{0.1000}& 0.9333\\

Lower Sepik-Ramu& \bf{0.2000}& 0.2500& \bf{0.2000}& 0.2500& 0.2500& \bf{0.2000}& 0.9923\\

Lakes Plain& \bf{0.4231}& \bf{0.4231}& \bf{0.4231}& \bf{0.4231}& \bf{0.4231}& \bf{0.4231}& 0.9929\\

Marind& \bf{0.3438}& \bf{0.3438}& \bf{0.3438}& \bf{0.3438}& \bf{0.3438}& \bf{0.3438}& 0.9947\\

Morehead and Upper Maro Rivers& \bf{0.3235}& \bf{0.3235}& \bf{0.3235}& \bf{0.3235}& \bf{0.3235}& \bf{0.3235}& 1.0000\\

Sepik& 0.4038& 0.3654& 0.3654& 0.3654& \bf{0.3269}& 0.3654& 1.0000\\

Sko& \bf{0.1429}& \bf{0.1429}& \bf{0.1429}& \bf{0.1429}& \bf{0.1429}& \bf{0.1429}& 0.9667\\

Australian& \bf{0.4113}& 0.4274& 0.4220& \bf{0.4113}& 0.4274& 0.4167& 1.0000\\

Torricelli& 0.4839& 0.4839& 0.4839& \bf{0.4516}& \bf{0.4516}& 0.4839& 1.0000\\

Trans-New Guinea& 0.4454& \bf{0.4386}& 0.4420& \bf{0.4386}& \bf{0.4386}& \bf{0.4386}& 0.9995\\

Western Fly& \bf{0.4615}& \bf{0.4615}& \bf{0.4615}& \bf{0.4615}& \bf{0.4615}& \bf{0.4615}& 1.0000\\

West Papuan& 0.4706& 0.4412& \bf{0.4118}& 0.4706& 0.4412& 0.4412& 1.0000\\

\hline

\multicolumn{8}{c}{ \bf AMERICA}\\ \hline

Na-Dene& 0.5455& 0.5000& \bf{0.4545}& 0.5455& 0.5000& \bf{0.4545}& 0.9933\\

Uto-Aztecan& 0.1914& 0.1914& 0.1914& 0.1914& \bf{0.1790}& 0.1914& 0.9959\\

Algic& 0.5000& 0.5000& 0.5000& \bf{0.4643}& 0.5000& 0.5000& 1.0000\\

Panoan& \bf{0.5000}& \bf{0.5000}& \bf{0.5000}& \bf{0.5000}& \bf{0.5000}& \bf{0.5000}& 1.0000\\

Salishan& \bf{0.2917}& \bf{0.2917}& \bf{0.2917}& \bf{0.2917}& \bf{0.2917}& \bf{0.2917}& 1.0000\\

Quechuan& \bf{0.4167}& \bf{0.4167}& \bf{0.4167}& \bf{0.4167}& \bf{0.4167}& \bf{0.4167}& 1.0000\\

Penutian& 0.2619& \bf{0.2143}& \bf{0.2143}& 0.2619& \bf{0.2143}& \bf{0.2143}& 0.9931\\

Tupian& \bf{0.4681}& \bf{0.4681}& \bf{0.4681}& \bf{0.4681}& \bf{0.4681}& \bf{0.4681}& 1.0000\\

Hokan& 0.4375& \bf{0.3958}& \bf{0.3958}& \bf{0.3958}& 0.4375& \bf{0.3958}& 1.0000\\

Macro-Ge& \bf{0.4583}& \bf{0.4583}& 0.5000& 0.5000& \bf{0.4583}& 0.5000& 1.0000\\

Oto-Manguean& \bf{0.3583}& \bf{0.3583}& \bf{0.3583}& \bf{0.3583}& \bf{0.3583}& \bf{0.3583}& 0.9945\\

Tucanoan& 0.4286& 0.4286& 0.4286& \bf{0.3571}& 0.4286& 0.4286& 0.9923\\

Arawakan& 0.4375& 0.4375& 0.4375& 0.4375& 0.4375& \bf{0.4167}& 1.0000\\

Cariban& 0.5263& 0.5263& \bf{0.4737}& 0.5263& 0.5789& 0.5263& 0.9833\\

Mixe-Zoque& \bf{0.1786}& 0.2500& 0.2500& \bf{0.1786}& 0.2500& 0.2500& 0.9867\\

Mayan& 0.4600& \bf{0.4067}& 0.4333& 0.4467& \bf{0.4067}& 0.4333& 1.0000\\

Chibchan& \bf{0.3500}& \bf{0.3500}& \bf{0.3500}& 0.4000& 0.4000& 0.4000& 0.9900\\

Totonacan& \bf{0.2143}& \bf{0.2143}& \bf{0.2143}& \bf{0.2143}& \bf{0.2143}& \bf{0.2143}& 1.0000\\

\hline

AVERAGE& 0.3998& 0.3970& \bf{0.3898}& 0.3963& 0.3962& 0.3910& 0.9951\\

\hline

\end{tabular}

}

\end{center}

\caption{ {\bf Robinson-Foulds distance between inferred trees and the Ethnologue classification.}  This table shows the Robinson-Foulds distances between the inferred trees and the Ethnologue classification for each language family in the ASJP database. For each language family we report the RF distance obtained using both the LDN (\ref{LDN2}) and the LDND (\ref{LDND}) distances between languages in the framework of the NJ, FastMe and FastSBiX algorithms. Finally the last column reports the results obtained in the comparison of the Ethnologue classifications with specific artificial random trees for each language family. First of all the RF distances for the actual inferred trees are much smaller than the corresponding random case. Still these values are always very high independently on the matrix distance used and of the specific phylogenetic algorithm adopted. A systematic effect that increases the errors is the bias induced by the non-binary nodes in the Ethnologue trees.} \label{RF_table} \end{table}

\begin{table}[htp]


\begin{center}\resizebox*{14cm}{!}{

\begin{tabular}{lccccccc}

\multicolumn{8}{c}{\bf GENERALIZED ROBINSON-FOULDS SCORE}\\\\ \hline

&\multicolumn{3}{|c|}{\bf LDN}&\multicolumn{3}{|c|}{\bf LDND}\\ \hline

&&\\

 &\bf Neighbour-Joining& \bf FastME& \bf FastSBiX& \bf Neighbour-Joining& \bf FastME& \bf FastSBiX& \bf RANDOM\\

&&\\

\hline

\multicolumn{8}{c}{ \bf AFRICA}\\ \hline

Khoisan& \bf{0.4615}& \bf{0.4615}& \bf{0.4615}& \bf{0.4615}& \bf{0.4615}& \bf{0.4615}& 0.9769\\

Niger-Congo& 0.4127& \bf{0.3891}& 0.4091& 0.4109& 0.4018& 0.4127& 0.9911\\

Nilo-Saharan& 0.3119& 0.3395& 0.2844& 0.3028& 0.2844& \bf{0.2752}& 0.9853\\

Kadugli& \bf{0.0000}& \bf{0.0000}& \bf{0.0000}& \bf{0.0000}& \bf{0.0000}& \bf{0.0000}& \bf{0.0000}\\

Afro-Asiatic& 0.2500& 0.2325& 0.2193& 0.2544& 0.2237& \bf{0.2149}& 0.9908\\

\hline

\multicolumn{8}{c}{ \bf EURASIA}\\ \hline

Indo-European& 0.3505& 0.3411& 0.3224& 0.3411& 0.3271& \bf{0.3084}& 0.9785\\

Uralic& \bf{0.1500}& \bf{0.1500}& \bf{0.1500}& \bf{0.1500}& \bf{0.1500}& \bf{0.1500}& 0.9450\\

Altaic& \bf{0.3108}& \bf{0.3108}& 0.3243& \bf{0.3108}& \bf{0.3108}& 0.3378& 0.9527\\

Dravidian& 0.6111& 0.6111& 0.4444& 0.5556& 0.5556& \bf{0.3889}& 0.9889\\

Nakh-Daghestanian& \bf{0.0690}& \bf{0.0690}& \bf{0.0690}& \bf{0.0690}& \bf{0.0690}& 0.1034& 0.9655\\

Sino-Tibetan& 0.4348& 0.4275& \bf{0.4130}& 0.4348& \bf{0.4130}& \bf{0.4130}& 0.9841\\

Hmong-Mien& \bf{0.1818}& \bf{0.1818}& 0.2727& 0.2727& \bf{0.1818}& 0.2727& 0.8636\\

Tai-Kadai& 0.4118& \bf{0.3922}& 0.4118& 0.4118& \bf{0.3922}& 0.4118& 0.9725\\

Great Andamanese& \bf{0.2857}& \bf{0.2857}& \bf{0.2857}& \bf{0.2857}& \bf{0.2857}& \bf{0.2857}& 0.4143\\

Austro-Asiatic& 0.3469& 0.3469& 0.3061& \bf{0.2245}& 0.2449& 0.3265& 0.9980\\

\hline

\multicolumn{8}{c}{ \bf PACIFIC }\\ \hline

Austronesian& 0.3881& 0.4063& 0.3820& 0.3844& 0.3942& \bf{0.3723}& 0.9793\\

Border& 0.0769& 0.0769& \bf{0.0000}& 0.0769& 0.0769& \bf{0.0000}& 0.7538\\

Bosavi& \bf{0.0000}& \bf{0.0000}& \bf{0.0000}& \bf{0.0000}& \bf{0.0000}& \bf{0.0000}& \bf{0.0000}\\

Kiwaian& \bf{0.0000}& \bf{0.0000}& \bf{0.0000}& \bf{0.0000}& \bf{0.0000}& \bf{0.0000}& \bf{0.0000}\\

Eleman& \bf{0.0000}& \bf{0.0000}& \bf{0.0000}& \bf{0.0000}& \bf{0.0000}& \bf{0.0000}& 0.8429\\

Lower Sepik-Ramu& \bf{0.0000}& 0.1176& \bf{0.0000}& 0.0588& 0.0588& \bf{0.0000}& 0.9765\\

Lakes Plain& \bf{0.1739}& \bf{0.1739}& \bf{0.1739}& \bf{0.1739}& \bf{0.1739}& \bf{0.1739}& 0.9043\\

Marind& \bf{0.0690}& \bf{0.0690}& \bf{0.0690}& \bf{0.0690}& \bf{0.0690}& \bf{0.0690}& 0.9517\\

Morehead and Upper Maro Rivers& 0.1429& 0.1429& 0.1429& \bf{0.0714}& 0.1429& \bf{0.0714}& 0.9500\\

Sepik& 0.0870& 0.0435& 0.0435& 0.0435& \bf{0.0000}& 0.0435& 0.9391\\

Sko& \bf{0.0000}& \bf{0.0000}& \bf{0.0000}& \bf{0.0000}& \bf{0.0000}& \bf{0.0000}& 0.5600\\

Australian& \bf{0.3653}& 0.4012& 0.3772& 0.3832& 0.3892& \bf{0.3653}& 0.9934\\

Torricelli& 0.2143& 0.2500& 0.2143& 0.2143& \bf{0.1786}& 0.2500& 0.9000\\

Trans-New Guinea& 0.2544& 0.2230& 0.2474& \bf{0.2195}& 0.2404& 0.2265& 0.9868\\

Western Fly& \bf{0.0000}& \bf{0.0000}& \bf{0.0000}& \bf{0.0000}& \bf{0.0000}& \bf{0.0000}& \bf{0.0000}\\

West Papuan& 0.1935& 0.1290& \bf{0.0968}& 0.2258& 0.1613& 0.1613& 0.9129\\

\hline

\multicolumn{8}{c}{ \bf AMERICA }\\ \hline

Na-Dene& 0.7368& 0.6316& \bf{0.5789}& 0.7368& 0.6316& \bf{0.5789}& 0.9842\\

Uto-Aztecan& 0.1622& 0.1622& 0.1622& 0.1622& \bf{0.1351}& 0.1892& 0.9405\\

Algic& 0.3846& 0.3846& \bf{0.3462}& 0.3846& 0.3846& \bf{0.3462}& 0.9385\\

Panoan& 0.8000& 0.8000& \bf{0.7333}& 0.8000& 0.8000& \bf{0.7333}& 0.9800\\

Salishan& \bf{0.1111}& \bf{0.1111}& \bf{0.1111}& \bf{0.1111}& \bf{0.1111}& \bf{0.1111}& 0.9556\\

Quechuan& \bf{0.0000}& \bf{0.0000}& \bf{0.0000}& \bf{0.0000}& \bf{0.0000}& \bf{0.0000}& \bf{0.0000}\\

Penutian& 0.1667& \bf{0.0556}& \bf{0.0556}& 0.1111& \bf{0.0556}& \bf{0.0556}& 0.9667\\

Tupian& 0.4444& 0.4000& 0.4222& \bf{0.3556}& 0.4667& 0.4667& 0.9867\\

Hokan& 0.4000& \bf{0.3500}& \bf{0.3500}& \bf{0.3500}& 0.4500& 0.4000& 0.9800\\

Macro-Ge& 0.3810& 0.3810& \bf{0.3333}& 0.3810& \bf{0.3333}& 0.3810& 0.9333\\

Oto-Manguean& \bf{0.0357}& \bf{0.0357}& \bf{0.0357}& \bf{0.0357}& \bf{0.0357}& \bf{0.0357}& 0.9393\\

Tucanoan& 0.1875& 0.1875& 0.2500& \bf{0.1250}& 0.1875& 0.1875& 0.9625\\

Arawakan& 0.2195& 0.1951& 0.1951& 0.2195& 0.1951& \bf{0.1707}& 0.9463\\

Cariban& 0.8125& \bf{0.7500}& \bf{0.7500}& \bf{0.7500}& 0.8750& 0.8750& 0.9500\\

Mixe-Zoque& \bf{0.1111}& 0.2222& 0.2222& \bf{0.1111}& 0.2222& 0.2222& 0.9889\\

Mayan& 0.1972& \bf{0.1268}& 0.1549& 0.1831& \bf{0.1268}& 0.1549& 0.9394\\

Chibchan& 0.4000& 0.4000& \bf{0.3333}& 0.5333& 0.5333& 0.5333& 0.9600\\

Totonacan& \bf{0.0000}& \bf{0.0000}& \bf{0.0000}& \bf{0.0000}& \bf{0.0000}& \bf{0.0000}& 0.7571\\

\hline

AVERAGE& 0.2470& 0.2401& \bf{0.2276}& 0.2399& 0.2394& 0.2354& 0.8320\\

\hline

\end{tabular}

}

\end{center}

 \caption{ {\bf Generalized Robinson-Foulds scores between inferred trees and the Ethnologue classification.}  This table shows the generalized Robinson-Foulds scores between the inferred trees and the Ethnologue classification. The legend is the same as in Table~\ref{RF_table}. The generalization has been introduced to get rid of biases in the Robinson-Foulds distance, due to the presence of non binary nodes in the classifications trees. The average scores are now much smaller (roughly for a factor $2$) than for the usual Robinson-Foulds distance.}  \label{GRF_table} \end{table}

\begin{table}[htp] 

\begin{center}\resizebox*{14cm}{!}{

\begin{tabular}{lccccccc}

\multicolumn{8}{c}{\bf QUARTET DISTANCE}\\\\ \hline

&\multicolumn{3}{|c|}{\bf LDN}&\multicolumn{3}{|c|}{\bf LDND}\\ \hline

&&\\

 &\bf Neighbour-Joining& \bf FastME& \bf FastSBiX& \bf Neighbour-Joining& \bf FastME& \bf FastSBiX& \bf RANDOM\\

&&\\

\hline

\multicolumn{8}{c}{ \bf AFRICA }\\ \hline

Khoisan& \bf{0.2885}& \bf{0.2885}& 0.2984& \bf{0.2885}& \bf{0.2885}& 0.2984& 0.7088\\

Niger-Congo& 0.3396& 0.5057& 0.3401& 0.3471& 0.3355& \bf{0.2853}& 0.6120\\

Nilo-Saharan& 0.3420& 0.3484& 0.3311& 0.2141& 0.2214& \bf{0.2128}& 0.6912\\

Kadugli& \bf{1.0000}& \bf{1.0000}& \bf{1.0000}& \bf{1.0000}& \bf{1.0000}& \bf{1.0000}& \bf{1.0000}\\

Afro-Asiatic& 0.2550& 0.2808& 0.2556& 0.2508& 0.2371& \bf{0.2083}& 0.6440\\

\hline

\multicolumn{8}{c}{ \bf EURASIA }\\ \hline

Indo-European& 0.1694& 0.1743& 0.1661& 0.1650& 0.1632& \bf{0.1626}& 0.6065\\

Uralic& 0.5007& \bf{0.4949}& \bf{0.4949}& 0.5007& \bf{0.4949}& 0.4963& 0.8087\\

Altaic& 0.2800& 0.2813& \bf{0.2733}& 0.2800& 0.2813& 0.2891& 0.7167\\

Dravidian& 0.3666& 0.3666& 0.3793& 0.3101& 0.3078& \bf{0.2526}& 0.7095\\

Nakh-Daghestanian& \bf{0.2613}& \bf{0.2613}& \bf{0.2613}& \bf{0.2613}& \bf{0.2613}& 0.2969& 0.7455\\

Sino-Tibetan& 0.5051& 0.4829& 0.4828& 0.4900& 0.4907& \bf{0.4821}& 0.7879\\

Hmong-Mien& \bf{0.4016}& \bf{0.4016}& 0.4066& 0.4066& \bf{0.4016}& 0.4066& 0.7690\\

Tai-Kadai& 0.3772& 0.3523& 0.3300& 0.3701& 0.3523& \bf{0.3245}& 0.7093\\

Great Andamanese& \bf{0.9571}& \bf{0.9571}& \bf{0.9571}& \bf{0.9571}& \bf{0.9571}& \bf{0.9571}& 0.9805\\

Austro-Asiatic& 0.3336& 0.2923& 0.2593& \bf{0.2549}& 0.2558& 0.2763& 0.7381\\

\hline

\multicolumn{8}{c}{ \bf PACIFIC }\\ \hline

Austronesian& 0.3731& 0.3721& 0.3306& 0.2963& 0.3976& \bf{0.2746}& 0.6650\\

Border& 0.6269& 0.6269& \bf{0.5692}& 0.6269& 0.6269& \bf{0.5692}& 0.8344\\

Bosavi& \bf{1.0000}& \bf{1.0000}& \bf{1.0000}& \bf{1.0000}& \bf{1.0000}& \bf{1.0000}& \bf{1.0000}\\

Kiwaian& \bf{1.0000}& \bf{1.0000}& \bf{1.0000}& \bf{1.0000}& \bf{1.0000}& \bf{1.0000}& \bf{1.0000}\\

Eleman& \bf{0.1190}& \bf{0.1190}& \bf{0.1190}& \bf{0.1190}& \bf{0.1190}& \bf{0.1190}& 0.6895\\

Lower Sepik-Ramu& \bf{0.1969}& 0.2140& \bf{0.1969}& 0.2004& 0.2004& \bf{0.1969}& 0.7354\\

Lakes Plain& \bf{0.4813}& \bf{0.4813}& \bf{0.4813}& \bf{0.4813}& \bf{0.4813}& \bf{0.4813}& 0.7817\\

Marind& \bf{0.2000}& \bf{0.2000}& \bf{0.2000}& \bf{0.2000}& \bf{0.2000}& \bf{0.2000}& 0.7345\\

Morehead and Upper Maro Rivers& \bf{0.2664}& 0.3496& \bf{0.2664}& 0.3345& 0.3546& 0.3345& 0.7282\\

Sepik& 0.3466& 0.3523& 0.3145& 0.3451& \bf{0.3130}& 0.3451& 0.7738\\

Sko& \bf{0.4857}& \bf{0.4857}& \bf{0.4857}& \bf{0.4857}& \bf{0.4857}& \bf{0.4857}& 0.7571\\

Australian& 0.5674& \bf{0.5553}& 0.5685& 0.5792& 0.5666& 0.5573& 0.7824\\

Torricelli& 0.4898& 0.4625& 0.4625& 0.4513& \bf{0.4493}& 0.4625& 0.8037\\

Trans-New Guinea& 0.4593& 0.4592& 0.4527& 0.4532& \bf{0.4474}& 0.4505& 0.7265\\

Western Fly& \bf{1.0000}& \bf{1.0000}& \bf{1.0000}& \bf{1.0000}& \bf{1.0000}& \bf{1.0000}& \bf{1.0000}\\

West Papuan& 0.2529& 0.2455& \bf{0.2209}& 0.2471& 0.2397& 0.2397& 0.7309\\

\hline

\multicolumn{8}{c}{ \bf AMERICA }\\ \hline

Na-Dene& 0.3794& 0.3753& \bf{0.3671}& 0.3794& 0.3753& \bf{0.3671}& 0.7234\\

Uto-Aztecan& \bf{0.2914}& \bf{0.2914}& \bf{0.2914}& \bf{0.2914}& 0.2945& 0.2965& 0.7419\\

Algic& \bf{0.5673}& \bf{0.5673}& 0.5695& 0.5885& 0.5941& 0.5782& 0.7717\\

Panoan& 0.6876& 0.6909& \bf{0.6791}& 0.6928& 0.6909& 0.6827& 0.8152\\

Salishan& \bf{0.3354}& \bf{0.3354}& \bf{0.3354}& \bf{0.3354}& \bf{0.3354}& \bf{0.3354}& 0.7578\\

Quechuan& \bf{1.0000}& \bf{1.0000}& \bf{1.0000}& \bf{1.0000}& \bf{1.0000}& \bf{1.0000}& \bf{1.0000}\\

Penutian& 0.2145& \bf{0.1370}& \bf{0.1370}& 0.2092& \bf{0.1370}& \bf{0.1370}& 0.7068\\

Tupian& 0.5359& 0.5188& 0.5109& \bf{0.5055}& 0.5135& 0.5209& 0.8169\\

Hokan& 0.2577& 0.2582& 0.2690& \bf{0.2447}& 0.2733& 0.2654& 0.7255\\

Macro-Ge& \bf{0.4343}& 0.4551& 0.4650& 0.4789& 0.4645& 0.4823& 0.7764\\

Oto-Manguean& \bf{0.3767}& \bf{0.3767}& \bf{0.3767}& \bf{0.3767}& \bf{0.3767}& \bf{0.3767}& 0.7855\\

Tucanoan& 0.2608& 0.2608& 0.3351& \bf{0.2567}& 0.2608& 0.2608& 0.7166\\

Arawakan& 0.3494& 0.3685& 0.3212& 0.3541& 0.3685& \bf{0.3133}& 0.7363\\

Cariban& 0.6370& 0.5986& 0.5929& \bf{0.5815}& 0.6391& 0.6373& 0.7304\\

Mixe-Zoque& \bf{0.2384}& 0.2566& 0.2566& \bf{0.2384}& 0.2566& 0.2566& 0.7402\\

Mayan& 0.2060& \bf{0.1925}& 0.1943& 0.2057& \bf{0.1925}& 0.1943& 0.7074\\

Chibchan& \bf{0.5788}& \bf{0.5788}& 0.6023& 0.6199& 0.6199& 0.6199& 0.8291\\

Totonacan& \bf{0.7000}& \bf{0.7000}& \bf{0.7000}& \bf{0.7000}& \bf{0.7000}& \bf{0.7000}& 0.9224\\

\hline

AVERAGE& 0.4550& 0.4566& 0.4471& 0.4485& 0.4494& \bf{0.4426}& 0.7744\\

\hline

\end{tabular}

}

\end{center}


\caption{ {\bf Quartet Distance between inferred trees and the Ethnologue classification.}  This table shows the Quartet Distance between the inferred trees and the Ethnologue classification for each language family in the ASJP database. The legend is the same as in Table~\ref{RF_table}. In this case the scores are affected by the presence of star quartets of taxa in the Ethnologue trees, as a consequence of the existence of non-binary nodes.}

\label{QD_table}

\end{table}

\begin{table}[htp] 


\begin{center}\resizebox*{14cm}{!}{

\begin{tabular}{lccccccc}

\multicolumn{8}{c}{\bf GENERALIZED QUARTET DISTANCE}\\\\ \hline

&\multicolumn{3}{|c|}{\bf LDN}&\multicolumn{3}{|c|}{\bf LDND}\\ \hline

&&\\

 &\bf Neighbour-Joining& \bf FastME& \bf FastSBiX& \bf Neighbour-Joining& \bf FastME& \bf FastSBiX& \bf RANDOM\\

&&\\

\hline

\multicolumn{8}{c}{ \bf AFRICA }\\ \hline

Khoisan& \bf{0.1809}& \bf{0.1809}& 0.1923& \bf{0.1809}& \bf{0.1809}& 0.1923& 0.6650\\

Niger-Congo& 0.1756& 0.3830& 0.1763& 0.1850& 0.1705& \bf{0.1078}& 0.5158\\

Nilo-Saharan& 0.2436& 0.2510& 0.2310& 0.0966& 0.1051& \bf{0.0951}& 0.6451\\

Kadugli& \bf{0.0000}& \bf{0.0000}& \bf{0.0000}& \bf{0.0000}& \bf{0.0000}& \bf{0.0000}& \bf{0.0000}\\

Afro-Asiatic& 0.0895& 0.1209& 0.0901& 0.0843& 0.0676& \bf{0.0323}& 0.5647\\

\hline

\multicolumn{8}{c}{ \bf EURASIA }\\ \hline

Indo-European& 0.0738& 0.0793& 0.0701& 0.0689& 0.0669& \bf{0.0662}& 0.5609\\

Uralic& 0.0458& \bf{0.0347}& \bf{0.0347}& 0.0458& \bf{0.0347}& 0.0373& 0.6355\\

Altaic& 0.0990& 0.1006& \bf{0.0906}& 0.0988& 0.1006& 0.1102& 0.6451\\

Dravidian& 0.2978& 0.2978& 0.3119& 0.2352& 0.2327& \bf{0.1715}& 0.6779\\

Nakh-Daghestanian& \bf{0.0179}& \bf{0.0179}& \bf{0.0179}& \bf{0.0179}& \bf{0.0179}& 0.0653& 0.6632\\

Sino-Tibetan& 0.1411& 0.1026& 0.1024& 0.1149& 0.1161& \bf{0.1012}& 0.6319\\

Hmong-Mien& \bf{0.1243}& \bf{0.1243}& 0.1316& 0.1316& \bf{0.1243}& 0.1316& 0.6620\\

Tai-Kadai& 0.2838& 0.2553& 0.2296& 0.2758& 0.2553& \bf{0.2232}& 0.6659\\

Great Andamanese& 0.6786& 0.6786& 0.6786& 0.6786& 0.6786& 0.6786& \bf{0.6283}\\

Austro-Asiatic& 0.1489& 0.0962& 0.0542& \bf{0.0485}& 0.0496& 0.0757& 0.6658\\

\hline

\multicolumn{8}{c}{ \bf PACIFIC }\\ \hline

Austronesian& 0.3006& 0.2995& 0.2523& 0.2149& 0.3279& \bf{0.1907}& 0.5132\\

Border& 0.1339& 0.1339& \bf{0.0000}& 0.1339& 0.1339& \bf{0.0000}& 0.6156\\

Bosavi& \bf{0.0000}& \bf{0.0000}& \bf{0.0000}& \bf{0.0000}& \bf{0.0000}& \bf{0.0000}& \bf{0.0000}\\

Kiwaian& \bf{0.0000}& \bf{0.0000}& \bf{0.0000}& \bf{0.0000}& \bf{0.0000}& \bf{0.0000}& \bf{0.0000}\\

Eleman& \bf{0.0000}& \bf{0.0000}& \bf{0.0000}& \bf{0.0000}& \bf{0.0000}& \bf{0.0000}& 0.5705\\

Lower Sepik-Ramu& \bf{0.0000}& 0.0213& \bf{0.0000}& 0.0044& 0.0044& \bf{0.0000}& 0.6717\\

Lakes Plain& \bf{0.2096}& \bf{0.2096}& \bf{0.2096}& \bf{0.2096}& \bf{0.2096}& \bf{0.2096}& 0.6675\\

Marind& \bf{0.0284}& \bf{0.0284}& \bf{0.0284}& \bf{0.0284}& \bf{0.0284}& \bf{0.0284}& 0.6763\\

Morehead and Upper Maro Rivers& \bf{0.0673}& 0.1732& \bf{0.0673}& 0.1540& 0.1797& 0.1540& 0.6549\\

Sepik& 0.0490& 0.0573& 0.0022& 0.0468& \bf{0.0000}& 0.0468& 0.6716\\

Sko& \bf{0.0000}& \bf{0.0000}& \bf{0.0000}& \bf{0.0000}& \bf{0.0000}& \bf{0.0000}& 0.2714\\

Australian& 0.2405& \bf{0.2195}& 0.2426& 0.2614& 0.2391& 0.2230& 0.6182\\

Torricelli& 0.1592& 0.1144& 0.1144& 0.0959& \bf{0.0926}& 0.1144& 0.6769\\

Trans-New Guinea& 0.1221& 0.1221& 0.1114& 0.1122& \bf{0.1029}& 0.1078& 0.5561\\

Western Fly& \bf{0.0000}& \bf{0.0000}& \bf{0.0000}& \bf{0.0000}& \bf{0.0000}& \bf{0.0000}& \bf{0.0000}\\

West Papuan& 0.0714& 0.0623& \bf{0.0317}& 0.0642& 0.0550& 0.0550& 0.6654\\

\hline

\multicolumn{8}{c}{ \bf AMERICA}\\ \hline

Na-Dene& 0.2251& 0.2200& \bf{0.2098}& 0.2251& 0.2200& \bf{0.2098}& 0.6548\\

Uto-Aztecan& \bf{0.0911}& \bf{0.0911}& \bf{0.0911}& \bf{0.0911}& 0.0951& 0.0976& 0.6686\\

Algic& \bf{0.3609}& \bf{0.3609}& 0.3641& 0.3920& 0.4003& 0.3768& 0.6628\\

Panoan& 0.4885& 0.4937& \bf{0.4746}& 0.4970& 0.4937& 0.4805& 0.6974\\

Salishan& \bf{0.0600}& \bf{0.0600}& \bf{0.0600}& \bf{0.0600}& \bf{0.0600}& \bf{0.0600}& 0.6577\\

Quechuan& \bf{0.0000}& \bf{0.0000}& \bf{0.0000}& \bf{0.0000}& \bf{0.0000}& \bf{0.0000}& \bf{0.0000}\\

Penutian& 0.0958& \bf{0.0066}& \bf{0.0066}& 0.0897& \bf{0.0066}& \bf{0.0066}& 0.6626\\

Tupian& 0.1783& 0.1478& 0.1340& \bf{0.1244}& 0.1386& 0.1517& 0.6757\\

Hokan& 0.1006& 0.1012& 0.1143& \bf{0.0848}& 0.1195& 0.1099& 0.6674\\

Macro-Ge& \bf{0.1598}& 0.1907& 0.2052& 0.2260& 0.2046& 0.2310& 0.6677\\

Oto-Manguean& \bf{0.0015}& \bf{0.0015}& \bf{0.0015}& \bf{0.0015}& \bf{0.0015}& \bf{0.0015}& 0.6857\\

Tucanoan& 0.1105& 0.1105& 0.2001& \bf{0.1056}& 0.1105& 0.1105& 0.6593\\

Arawakan& 0.1859& 0.2098& 0.1506& 0.1916& 0.2098& \bf{0.1407}& 0.6702\\

Cariban& 0.5455& 0.4973& 0.4902& \bf{0.4761}& 0.5480& 0.5458& 0.6625\\

Mixe-Zoque& \bf{0.0528}& 0.0754& 0.0754& \bf{0.0528}& 0.0754& 0.0754& 0.6772\\

Mayan& 0.0469& \bf{0.0307}& 0.0327& 0.0465& \bf{0.0307}& 0.0327& 0.6483\\

Chibchan& \bf{0.1673}& \bf{0.1673}& 0.2140& 0.2488& 0.2488& 0.2488& 0.6625\\

Totonacan& \bf{0.0000}& \bf{0.0000}& \bf{0.0000}& \bf{0.0000}& \bf{0.0000}& \bf{0.0000}& 0.2224\\

\hline

AVERAGE& 0.0984& 0.0965& 0.0870& 0.0884& 0.0894& \bf{0.0825}& 0.4122\\

\hline

\end{tabular}

}

\end{center}

\caption{ {\bf Generalized Quartet Distance between inferred trees and the Ethnologue classification.} The generalized QD, shown in this table, quantifies the overall disagreement between inferred trees and the Ethnologue classifications. The legend is the same as in Table~\ref{RF_table}. The errors are now extremely low being always lower than $10\%$. This means that distance based approaches lead to accurate and robust classifications of the languages taken in account.}

\label{GQD_table}

\end{table}

\begin{table}[htp] 


\begin{center}\resizebox*{\textwidth}{!}{

\begin{tabular}{|c|c|c|c|c|}

\hline

Acehnese& Gapapaiwa& Kwara'ae (Solomon Islands')& Muyuw& Solos\\

\hline

Aklanon-Bisayan& Gaddang& Lahanan& Nalik& Soboyo\\

\hline

Alune& Gayo& Lala& Nanggu& Sowa\\

\hline

Amahai& Gedaged& Lamalera (lembata)& Nauna& Suau\\

\hline

Amara& Geser& Lamboya& Nehan& Sudest\\

\hline

Ambai (Yapen)& Ghari& Lamogai (Mulakaino)& Nengone& Surigaonon\\

\hline

Anakalang& Gorontalo (Hulondalo)& Lampung& Nggao (Poro)& Tabar\\

\hline

Apma Suru Kavian& Gumawana& Langalanga& Nggela& Tagabili\\

\hline

Aputai& Haku& Lau& Nila& Tagalog\\

\hline

Araki (Southwest Santo)& Hawaiian& Leipon& Niue& Tagbanwa, Aborlan Dialect\\

\hline

Arosi (Tawatana Village)& Hiligaynon& Lengo& Nukuoro& Tagbanwa, Kalamian, Coron Island Dialect\\

\hline

As& Hitu (Ambon)& Letinese& Numfor& Tahitian (Modern)\\

\hline

Asumboa& Hiw& Levei& Ogan& Taiof\\

\hline

Bali& Hoava& Likum& Oroha& Takia\\

\hline

Banggai (W.dialect)& Iaai& Lio, Flores Tongah& Paiwan& Talur\\

\hline

Banoni& Iban& Longgu& Palauan& Tanga\\

\hline

Bantik& Ibanag& Loniu& Palu'e (Nitung)& Tarpia\\

\hline

Baree& Idaan& Lou& Pangasinan& Tausug, Jolo Dialect\\

\hline

Barok& Iliun& Luang& Papora& Tawala\\

\hline

Bauro (Baroo Village)& Ilokano& Luangiua& Patpatar& Teanu\\

\hline

Belait& Ilongot;Kakiduge:n& Lunga Lunga (Minigir)& Paulohi& Tela-Masbuar\\

\hline

Besemah& Imorod& Lungga& Pazeh& Teop\\

\hline

Biga (Misool)& Imroing& Maanyan& Penrhyn& Teun\\

\hline

Bilur& Inabaknon& Madak& Perai& Thao\\

\hline

Bima& Indonesian& Madurese& Phan Rang Cham (Eastern Cham)& Tiang\\

\hline

Bintulu& Inibaloi& Magori (South East Papua)& Pukapuka& Tigak\\

\hline

Binukid& Iranun& Maisin& Pulo-Annan& Tikopia\\

\hline

Blablanga& Itneg, Binongan& Malango& Puluwatese& Timugon (Murut)\\

\hline

Bobot& Ivatan, Basco Dialect& Maleu& Puyuma& Tolo\\

\hline

Bolaang Mongondow& Jawe& Mamanwa& Raga& Tongan\\

\hline

Bonerate& Kadorih& Mamboru& Rarotongan& Tonsea\\

\hline

Bonfia& Kahua& Manam& Rejang Rejang& Tontemboan\\

\hline

Bughotu& Kaidipang& Manggarai& Rennellese& Torau\\

\hline

Bukat& Kairiru& Manihiki& Ririo& Tsou\\

\hline

Buli& Kalagan& Manobo, Ata (down river)& Roma& Tugun\\

\hline

Bunun& Kalinga, Limos& Manobo, Dibabawon& Roro& Tunjung\\

\hline

Buol& Kallahan, Keleyqiq& Manobo, Ilianen (Kibudtungan Dialect)& Rotuman& Ubir\\

\hline

Butuanon& Kambera& Manobo, Sarangani, Kayaponga Dialect& Roviana& Ughele\\

\hline

Carolinian& Kanakanabu& Mansaka& Rukai& Ujir (N.Aru)\\

\hline

Cebuano& Kandas& Maori& Rurutuan& Ura\\

\hline

Centra Amis& Kapampangan& Mapun& Sa (south easterndialect)& Uruava\\

\hline

Chamorro& Kapingamarangi& Maranao& Sa'a& Vaeakau-Taumako\\

\hline

Chru& Katingan& Marau& Saaroa& Vaghua\\

\hline

Chuukese (AKA Trukese)& Kaulong (Au Village) & Marovo& Sambal, Botolan& Varisi\\

\hline

Dai& Kavalan& Marshallese& Samoan& Vitu\\

\hline

Dawera-Daweloor& Kayupulau[Kajupulau]& Masiwang& Sangir& Wampar\\

\hline

Dehu& Kazukuru& Matukar& Sasak& Wanukaka\\

\hline

Diodio& Kemak& Maututu& Savu& Waray-Waray\\

\hline

Dorio& Kerinci& Mbaelelea& Seimat& Waropen\\

\hline

Doura& Kilivila& Mbaengguu& Sekar& Watubela\\

\hline

Emae& Kiribati& Mbirao& Selaru& Wedau\\

\hline

Emplawas& Kis& Mekeo& Sengseng& Windesi Wandamen\\

\hline

Ende& Kisar& Mengen& Serili& Wogeo\\

\hline

Erai& Kokota& Minangkabau& Serua& Woleaian\\

\hline

Fagani& Komering& Modang& Siar& Wolio\\

\hline

Fataleka& Kove& Moken& Sika& Wuvulu\\

\hline

Favorlang& Kuni& Molima& Simbo& Yabem\\

\hline

Fijian (Bau)& Kusaghe& Mono& Singhi& Yakan\\

\hline

Futuna-Aniwa& Kusaie& Mota& Siraya& Yamdena\\

\hline

Futuna, East& Kwai& Motu& Soa& Yapese\\

\hline

Gabadi& Kwaio& Nakanai (Bileki Dialect)& Sobei& Zabana (Kia)\\

\hline

\end{tabular}

}

\end{center}

\caption{ {\bf List of the 305 Austronesian languages considered in both ABVD and ASJP databases.} Here we report the complete list of the 305 languages taken in account in our studies of ABVD and ASJP list of words for the Austronesian family. Names of languages are reported as presented in  \bf{http://language.psy.auckland.ac.nz/austronesian/}. }

\label{ABVD_list}

\end{table}

\begin{table}[htp]


\begin{center}{

\begin{tabular}{lcccccc}

\hline

|&\multicolumn{3}{|c|}{\bf ABVD}&\multicolumn{3}{|c|}{\bf ASJP}\\ 

\hline

&\bf Neighbour-Joining& \bf FastME& \bf FastSBiX& \bf Neighbour-Joining& \bf FastME& \bf FastSBiX\\

\hline

RF& 0.5762& \bf{0.5573}& 0.6240& 0.6026& 0.5993& 0.5993\\

QD& 0.3508& 0.3657& \bf{0.2371}& 0.3605& 0.4594& 0.3163\\

GRF& 0.5464& \bf{0.5331}& 0.5430& 0.6192& 0.6325& 0.5927\\

GQD& 0.2860& 0.3024& \bf{0.1609}& 0.2969& 0.4055& 0.2815\\

\hline

\end{tabular}

}

\end{center}

\caption{ {\bf Accuracy in reconstructing the Austronesian family tree with the ABVD and ASJP database.} This table shows the Robinson-Foulds distance, the Quartet Distance and their generalizations, between the Austronesian language tree (inferred with both the ASJP and the ABVD database) and its relative Ethnologue-Classification for the languages shown in Table.\ref{ABVD_list}. In this case we only used the LDN definition of distance between lists of words and FastSBiX as distance-based algorithm to infer the trees. All the measures indicate that the ABVD database allows for a more accurate reconstruction of this language tree.}

\label{ABVD_accuracy}

\end{table}

\begin{figure}[htp] \begin{center} 

\includegraphics[width=\textwidth]{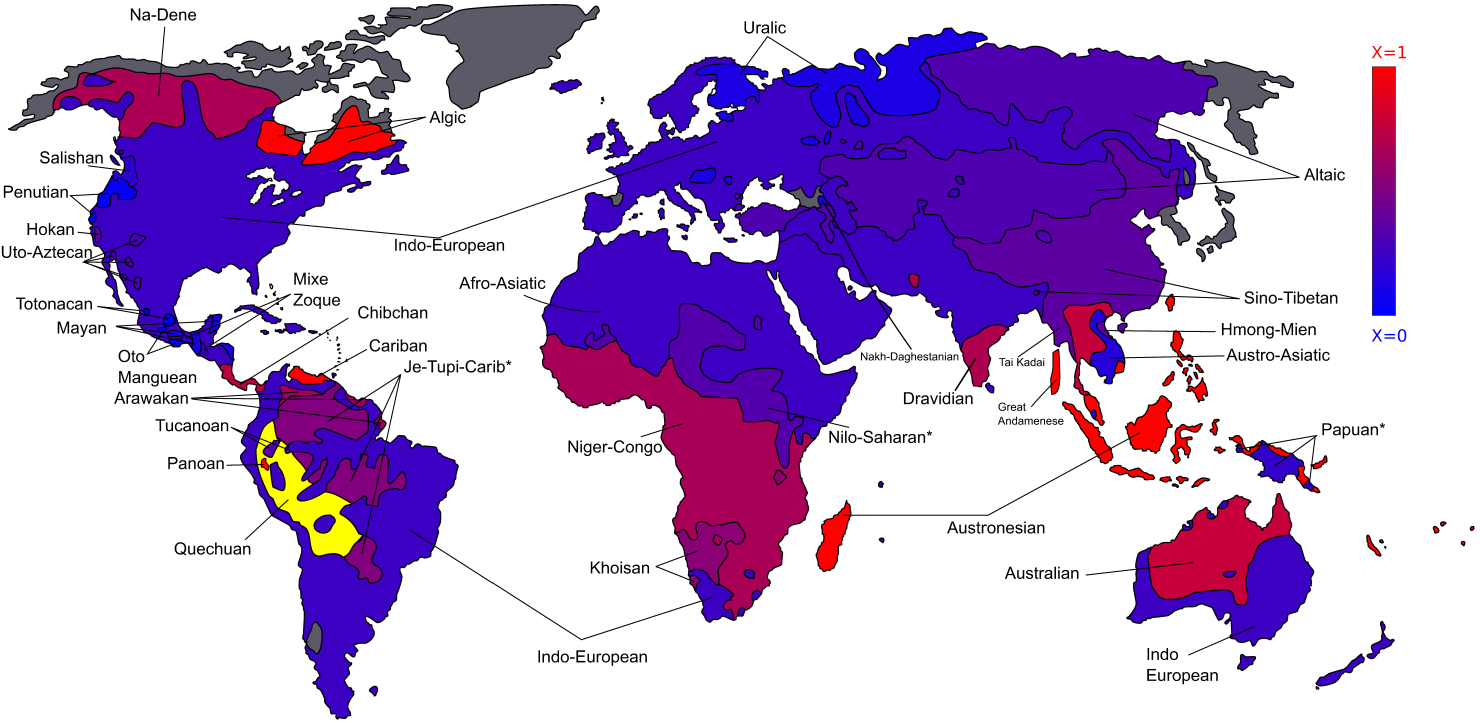}

\end{center} 

\caption{ {\bf Accuracy of the reconstruction across the planet. FastME, LDND matrix.} This map represents the level of accuracy of the FastME algorithm on several language families throughout the world. Trees Inferred with LDND matrices. The legend is the same of Fig.\ref{map_FastSBiX_LDND}.}

\label{map_FastME_LDND}

\end{figure}

\end{document}